\documentclass[twocolumn,twocolappendix]{aastex631}

\newcommand\starname{L\,98-59}
\newcommand\planetname{L\,98-59\,b}

\usepackage{times}
\usepackage{graphicx}%
\usepackage{multirow}%
\usepackage{amsmath,amssymb,amsfonts}%
\usepackage{mathrsfs}%
\usepackage[title]{appendix}%
\usepackage{xcolor}%
\usepackage{textcomp}%
\usepackage{booktabs}%
\usepackage{algorithm}%
\usepackage{algorithmicx}%
\usepackage{algpseudocode}%
\usepackage{listings}%
\usepackage{makecell}
\usepackage[version=3]{mhchem} % Formula subscripts using \ce{}

\shorttitle{A volcanic atmosphere on \planetname}
\shortauthors{Bello-Arufe et al.}
\begin{document}

\title{Evidence for a volcanic atmosphere on the sub-Earth L\,98-59\,b}

\correspondingauthor{Aaron Bello-Arufe}
\email{aaron.bello.arufe@jpl.nasa.gov}

\author[0000-0003-3355-1223]{Aaron Bello-Arufe}
\affiliation{Jet Propulsion Laboratory, California Institute of Technology, Pasadena, CA 91109, USA}

\author[0000-0002-1830-8260]{Mario Damiano}
\affiliation{Jet Propulsion Laboratory, California Institute of Technology, Pasadena, CA 91109, USA}

\author[0000-0002-9030-0132]{Katherine A. Bennett}
\affiliation{Department of Earth \& Planetary Sciences, Johns Hopkins University, Baltimore, MD 21218, USA}

\author[0000-0003-2215-8485]{Renyu Hu}
\affiliation{Jet Propulsion Laboratory, California Institute of Technology, Pasadena, CA 91109, USA}
\affiliation{Division of Geological and Planetary Sciences, California Institute of Technology, Pasadena, CA 91125, USA}

\author[0000-0003-0156-4564]{Luis Welbanks}
\altaffiliation{51 Pegasi b Fellow}
\affiliation{School of Earth and Space Exploration, Arizona State University, 781 Terrace Mall, Tempe, AZ 85287, USA}

\author[0000-0003-4816-3469]{Ryan J. MacDonald}
\altaffiliation{NHFP Sagan Fellow}
\affiliation{Department of Astronomy, University of Michigan, 1085 S. University Ave., Ann Arbor, MI 48109, USA}

\author[0000-0002-0726-6480]{Darryl Z. Seligman}
\altaffiliation{NSF Astronomy and Astrophysics Postdoctoral Fellow}
\affiliation{Department of Physics and Astronomy, Michigan State University, East Lansing, 48824,
MI, USA}

\author[0000-0001-6050-7645]{David K. Sing}
\affiliation{Department of Earth \& Planetary Sciences, Johns Hopkins University, Baltimore, MD 21218, USA}
\affiliation{Department of Physics \& Astronomy, Johns Hopkins University, Baltimore, MD 21218 USA}

\author[0000-0002-4675-9069]{Armen Tokadjian}
\affiliation{Jet Propulsion Laboratory, California Institute of Technology, Pasadena, CA 91109, USA}

\author[0000-0002-1655-0715]{Apurva Oza}
\affiliation{Division of Geological and Planetary Sciences, California Institute of Technology, Pasadena, CA 91125, USA}
\affiliation{Jet Propulsion Laboratory, California Institute of Technology, Pasadena, CA 91109, USA}

\author[0000-0002-1551-2610]{Jeehyun Yang}
\affiliation{Jet Propulsion Laboratory, California Institute of Technology, Pasadena, CA 91109, USA}

%% Note that the \and command from previous versions of AASTeX is now
%% depreciated in this version as it is no longer necessary. AASTeX 
%% automatically takes care of all commas and "and"s between authors names.

%% AASTeX 6.31 has the new \collaboration and \nocollaboration commands to
%% provide the collaboration status of a group of authors. These commands 
%% can be used either before or after the list of corresponding authors. The
%% argument for \collaboration is the collaboration identifier. Authors are
%% encouraged to surround collaboration identifiers with ()s. The 
%% \nocollaboration command takes no argument and exists to indicate that
%% the nearby authors are not part of surrounding collaborations.

%% Mark off the abstract in the ``abstract'' environment. 
\begin{abstract}

Assessing the prevalence of atmospheres on rocky planets around M-dwarf stars is a top priority of exoplanet science. High-energy activity from M-dwarfs can destroy the atmospheres of these planets, which could explain the lack of atmosphere detections to date. Volcanic outgassing has been proposed as a mechanism to replenish the atmospheres of tidally-heated rocky planets. \planetname, a sub-Earth transiting a nearby M dwarf, was recently identified as the most promising exoplanet to detect a volcanic atmosphere. We present the transmission spectrum of \planetname\ from four transits observed with JWST NIRSpec G395H. Although the airless model provides an adequate fit to the data based on its $\chi^2$, an \ce{SO2} atmosphere is preferred by 3.6$\sigma$ over a flat line in terms of the Bayesian evidence. Such an atmosphere would likely be in a steady state where volcanism balances escape. If so, \planetname\ must experience at least eight times as much volcanism and tidal heating per unit mass as Io. If volcanism is driven by runaway melting of the mantle, we predict the existence of a subsurface magma ocean in \planetname\ extending up to $R_p\sim 60-90\%$. An \ce{SO2}-rich volcanic atmosphere on \planetname\ would be indicative of an oxidized mantle with an oxygen fugacity of $f\rm{O}_2>IW+2.7$, and it would imply that \planetname\ must have retained some of its volatile endowment despite its proximity to its star. Our findings suggest that volcanism may revive secondary atmospheres on tidally heated rocky planets around M-dwarfs.

\end{abstract}

%% Keywords should appear after the \end{abstract} command. 
%% The AAS Journals now uses Unified Astronomy Thesaurus concepts:
%% https://astrothesaurus.org
%% You will be asked to selected these concepts during the submission process
%% but this old "keyword" functionality is maintained in case authors want
%% to include these concepts in their preprints.
\keywords{Exoplanet atmospheric composition (2021) --- Extrasolar rocky planets (511) --- James Webb Space Telescope (2291) --- M dwarf stars (982) --- Planetary interior (1248) --- Transmission spectroscopy (2133) --- Volcanism (2174)}

%% From the front matter, we move on to the body of the paper.
%% Sections are demarcated by \section and \subsection, respectively.
%% Observe the use of the LaTeX \label
%% command after the \subsection to give a symbolic KEY to the
%% subsection for cross-referencing in a \ref command.
%% You can use LaTeX's \ref and \label commands to keep track of
%% cross-references to sections, equations, tables, and figures.
%% That way, if you change the order of any elements, LaTeX will
%% automatically renumber them.
%%
%% We recommend that authors also use the natbib \citep
%% and \citet commands to identify citations.  The citations are
%% tied to the reference list via symbolic KEYs. The KEY corresponds
%% to the KEY in the \bibitem in the reference list below. 

\section{Introduction} \label{sec:intro}
The saga of scientific discovery often unfolds with thrilling anticipation, as exemplified by the historic encounter of Voyager 1 with Io in 1979. Just days before the flyby, \citet{Peale1979} published a seminal paper positing that Io's eccentric orbit would drive sufficient tidal dissipation to trigger a runaway melting process of its interior, foreshadowing the spectacular geological activity to come. Within days, Voyager 1 revealed the first-ever volcanic plumes beyond Earth, confirming the predictions and forever changing our understanding of planetary geology \citep{Smith1979,Morabito1979,Strom1979}.

More recently, the detectability of volcanism in planets and satellites beyond our Solar System has been hypothesized \citep[e.g.][]{kaltenegger2010,hu2013photochemistry,oza2019,Quick2020,ostberg2023,seligman2024tidal}. Volcanic activity is driven by both endogenous and/or exogenic processes. For example, terrestrial volcanism is driven largely by endogenous processes such as radiogenic heating. However, exogenic processes --- such as tidal dissipation ---  can dominate over endogenous ones in particular orbital and hierarchical configurations. While tidal heating in the Earth-moon system is relatively weak \citep{Peale1978}, shorter period and more eccentric exoplanets may experience sufficient tidal heating to drive widespread volcanic activity, as in the case of Io.

The sub-Earth-sized planet \planetname\ 
\citep[$R_p = 0.85~R_\earth$,][]{demangeon2021l9859b} has been identified as one of the most promising candidates for detecting active volcanism \citep{Quick2020,seligman2024tidal}. By generalizing the runaway melting mechanism from \citet{Peale1978} and \citet{Peale1979} to extrasolar planets, \citet{seligman2024tidal} estimate that tidal heating from its eccentric orbit ($e= 0.103^{+0.117}_{-0.045}$, \citealt{demangeon2021l9859b}, $e= 0.167^{+0.034}_{-0.16}$, \citealt{rajpaul2024}) could raise the equilibrium temperature of \planetname\ from $T_{\rm eq}\sim 600$~K to $\sim 1000$~K and drive widespread surface volcanism. Not only does \planetname\ have one of the largest predicted tidal heating rates, but it is also one of the most observable rocky exoplanets. Located only 10.6~pc away from the Earth, \planetname\ transits a bright M-type star ($m_{\rm J}=7.9$, $T_{\rm eff}=3415$~K) every 2.25 days. Its favorable planet-to-star radius ratio, bright host star, frequent transits and location at the border of the JWST continuous viewing zone make \planetname\ one of the most accessible terrestrial exoplanets in transmission spectroscopy with JWST. 

While high-energy activity from M-dwarfs can strip away the atmospheres of rocky planets around them, volcanic outgassing has been proposed as a mechanism to replenish them \citep[e.g.][]{kite2020}. The launch of JWST has finally enabled the search for high mean molecular weight atmospheres on rocky exoplanets around M dwarfs, but observations so far have either ruled out thick atmospheres or led to inconclusive results \citep[e.g.][]{lustigyaeger2023,greene2023,zieba2023,may2023gj1132b,moran2023gj486b,alderson2024,zhang2024,mansfield2024,wachiraphan2024,august2024,ducrot2024}.

JWST observations of the other two transiting planets in the \starname\ system, planets c and d (with radii of $1.34\pm0.07~R_\earth$ and $1.58\pm0.08~R_\earth$, respectively, \citealt{luquepalle2022}), were recently published. \citet{scarsdale2024} ruled out pure \ce{CH4} and $\lesssim 300\times$-solar metallicity atmospheres on the super-Earth \starname~c, while \citet{Gressier2024} and \citet{Banerjee2024} found hints of a sulfur atmosphere on \starname~d. Unlike the two innermost planets, planet d has a density inconsistent with that of a purely rocky planet \citep{demangeon2021l9859b,luquepalle2022} and should have a massive \ce{H2}/He/\ce{H2O} envelope, in which the formation of \ce{SO2} can be naturally expected from atmospheric thermochemistry and photochemistry \citep[e.g.][]{Yang_2024}. Here we present the transmission spectra of \planetname\ to search for evidence of a volcanic atmosphere. These constitute the first observations of \planetname\ with JWST. Previous Hubble observations of this planet have ruled out cloud-free hydrogen-dominated atmospheres \citep{damiano2022l9859b,zhou2022}, but high molecular weight atmospheres remain consistent with the data.

\section{Observations and Data Reduction} \label{sec:obs}
As part of JWST's Cycle 2 General Observers program \#3942 (PI: Damiano), we observed four transits of \planetname\ using JWST's Near InfraRed Spectrograph  \citep[NIRSpec,][]{jakobsen2022,birkmann2022}. We used the G395H grating, which disperses photons across two detectors, NRS1 and NRS2. This instrument mode exhibits a resolving power of $R = \lambda/\Delta\lambda \sim 2700$, where $\lambda$ is wavelength, and provides continuous spectral coverage from $2.7-5.2~\mu$m, except for a $\Delta \lambda = 0.1~\mu$m gap centered around 3.77~$\mu$m that separates the two detectors.

The observations took place on Jan 30, Feb 3, Feb 6, and Feb 19 2024 UTC, with each of them covering the full 0.97-hour transit and 1.87 hours of out-of-transit baseline to precisely measure the transit depth and model instrumental systematic noise. We used the Bright Object Time Series (BOTS) mode with the NRSRAPID readout pattern, the 1.6'' $\times$ 1.6'' fixed slit aperture (S1600A1), and the SUB2048 subarray. We obtained 3 groups per integration and 2822 integrations per time-series observation.

We extracted the transmission spectra of \planetname\ from the four transits using two independent pipelines, \texttt{Eureka!} \citep{bell2022eureka} and \texttt{FIREFLy} \citep{rustamkulov2022firefly,rustamkulov2023wasp39b}, to ensure that our results are robust against different data reduction methods. In Appendix~\ref{app:comparison}, we compare the transmission spectra from both reductions. The spectra are in good agreement across all transits and detectors.

\subsection{\texttt{Eureka!}}
We reduced the data using version 0.10 of the \texttt{Eureka!} pipeline \citep{bell2022eureka}. Starting from the uncalibrated raw files, and after testing different setups, we decided to run all the default steps in \texttt{Eureka!}'s stage 1, but we increased the jump rejection threshold from 4$\sigma$ to 7$\sigma$ due because it resulted in less noisy lightcurves. Before fitting the ramps, we performed a background subtraction at the group level using the mean value of the 32 outermost pixels (16 on each side) in each detector column. In the calculation of the mean background flux, we kept the spectral trace masked to avoid self-subtraction of the signal. The masked region included 9 pixels above and below the center of the trace.
In stage 2, we skipped the flat field and photometric calibration steps and ran all the other default steps.

\begin{figure*}
\centering
\includegraphics[width=\textwidth]{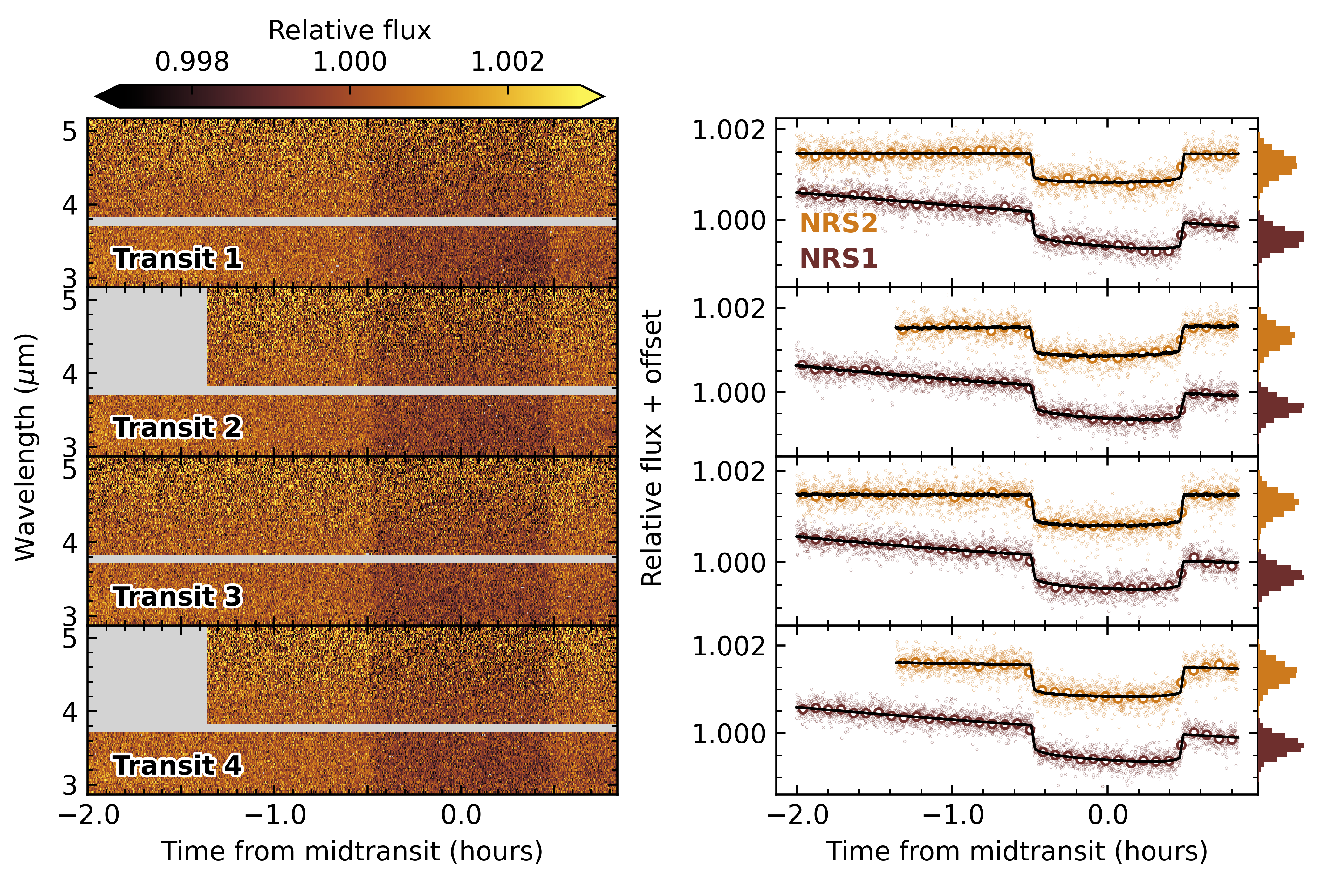}
\caption{\textit{Left:} Raw spectroscopic lightcurves, as extracted with \texttt{Eureka!} and binned to $\Delta\lambda = 0.02~\mu$m. The gray areas mark the separations between the data from the NRS1 and NRS2 detectors, as well as the integrations that were trimmed out of the NRS2 lightcurves of transits 2 and 4. \textit{Right:} White lightcurves and best-fit models. We also show the lightcurve data points binned by a factor of 80 to more easily identify the small undulations. On the right axes, we show the histograms of the unbinned residuals.}\label{fig1}
\end{figure*}

After the initial processing and calibration stages, we ran stages 3 and 4 of \texttt{Eureka!} to generate the lightcurves, shown in Fig.~\ref{fig1}.
We extracted columns 545--2041 in the NRS1 detector and 6--2044 in NRS2, and we masked all pixels with an odd (i.e., not even) data quality value. We straightened the trace and applied an additional round of background subtraction at the integration level. We then performed optimal extraction using the pixels within 3 rows from the center of the trace. During optimal extraction, we used the median integration as the spatial profile, after smoothing along the spectral direction with a 13-pixel long boxcar filter. Pixels that deviated by more than 10$\sigma$ from the spatial profile were rejected during optimal extraction. We computed the white lightcurves by binning the data between 2.87--3.71 $\mu$m in the case of NRS1, and 3.83--5.15~$\mu$m in the case of NRS2. We generated the spectroscopic lightcurves using bins with a width of $\Delta \lambda = 0.02~\mu$m, but we also tested bin widths of 0.01 and $0.04~\mu$m to explore the effect of spectral resolution on our inferences. We cleaned the lightcurves by removing 4$\sigma$ outliers using a boxcar filter with a width of 20 integrations. Low-frequency undulations in flux are apparent in the lightcurves (see also Fig.~\ref{fig:allan}). The time-correlated noise operating on timescales of $\sim 0.3-5$ minutes can be explained by thermal cycling of the heaters \citep{rigby2023}. However, a subset of the lightcurves additionally show lower-frequency flux undulations. We attribute these to the low number of groups per integration \citep[see e.g.][]{alderson2024,hu2024,wallack2024}. We trimmed the first 640 integrations of the second and fourth NRS2 observations, as they showed the most significant low-frequency undulations in flux.

We fit the white lightcurves with \texttt{emcee} \citep{foremanmackey2013emcee}, using a combination of a \texttt{batman} transit model \citep{kreidberg2015batman} and a systematics model. The systematics model included a polynomial in time and a linear decorrelation against drift in the spatial direction. We set the degree of the polynomial to 2 and 1 for the NRS1 and NRS2 lightcurves, respectively. We also fit a white noise multiplier to boost the uncertainties of the data points according to the scatter of the residuals. We first fitted each white lightcurve independently. We assumed a circular orbit and kept the orbital period fixed to 2.2531136 days \citep{demangeon2021l9859b}. We assigned a broad uniform prior to the midtransit time and imposed Gaussian priors on the orbital inclination ($i_p$) and scaled semi-major axis ($a/R_\star$) of $\mathcal{N}(87.71^\circ,4.00^\circ)$ and $\mathcal{N}(15,2)$, respectively \citep{demangeon2021l9859b}. We assigned a broad Gaussian prior to the planet-to-star radius ratio, and we fixed the quadratic limb-darkening coefficients to the values calculated with the \texttt{ExoTiC-LD} \citep{grantwakeford2022exoticld} package using 3D stellar models \citep{magic2015stagger} that assumed the stellar parameters reported in the literature: $T_{\rm eff}=3415$~K, $\log g = 4.86$, and $\lbrack \rm{Fe/H} \rbrack = -0.46$ \citep{demangeon2021l9859b}. We produced an additional \texttt{Eureka!} reduction in order to  ensure that our assumed limb-darkening coefficients are consistent with the data. In this reduction, we fitted for the quadratic limb-darkening coefficients ($q_1,q_2$, \citealt{kipping2013limbdark}). Specifically, we assigned flat priors from 0 to 1 to both $q_1$ and $q_2$. The derived values are consistent with those from \texttt{ExoTiC-LD}, as shown in Fig.~\ref{fig:ld_comparison}. The derived $q_1$ values center around 0.1, while $q_2$ has flat posteriors between 0 and 1 at all wavelengths. Freeing the limb-darkening coefficients produces a transmission spectrum that is consistent with our standard reduction (see Fig.~\ref{transmission_spectra_comparison_eurekaonly}).

We fit the spectroscopic lightcurves in a similar fashion to the white lightcurves, but we kept $i_p$ and $a/R_\star$ fixed to the weighted average of the best-fit parameters from all the white lightcurve fits. We also fixed the value of the mid-transit time, but in this case, the weighted average was taken on a transit-by-transit basis to avoid being biased by the potential presence of small transit timing variations \citep{cloutier2019l9859}. The quadratic term ($c_2$) of the polynomial in the NRS1 systematics model was fixed to the value obtained in the corresponding white lightcurve fit, but we kept a wavelength-dependent linear term for both NRS1 and NRS2 \citep{moran2023gj486b}. We also tried keeping $c_2$ free in the spectroscopic lightcurve fits, and the resulting transmission spectra did not significantly change (see Fig.~\ref{transmission_spectra_comparison_eurekaonly}). The Allan deviation plots and the best-fit orbital parameters from the fits are shown in Appendices~\ref{app:allan} and ~\ref{app:bestfit_values}, respectively.

\subsection{FIREFLy}\label{subsubsec:firefly_reduction}

We conduct a second reduction using the \texttt{FIREFLy} pipeline \citep{rustamkulov2022firefly, rustamkulov2023wasp39b}. This pipeline begins by running Stages 1 and 2 of the \texttt{jwst} reduction pipeline, which applies the standard group- and integration-level corrections. The only steps that \texttt{FIREFLy} changes at Stages 1 and 2 are 1) the addition of a group-level 1/f subtraction, and 2) the skipping of the dark, flat-field and jump correction steps. We do not apply the scaled superbias step discussed in \citet{moran2023gj486b}, as updated JWST calibration files have improved the automated superbias correction step in Stage 1. 

Following the integration-level Stage 2 instrument corrections, we run the data through our custom \texttt{FIREFLy} pipeline. We begin by applying cosmic ray cleaning using \texttt{lacosmic} \citep{vandokkum2001lacosmic}, while also manually examining a known bad G395H pixel. Another 1/f correction is applied at the integration-level, after which we measure the intrapixel shifts in the x- (spectral) and y- (spatial) directions, which may be used in the systematics model. The trace is measured using a fourth order polynomial, and from this, we extract the spectra with an aperture size of 4.8 pixels and 5.3 pixels (full-width) for NRS1 and NRS2, respectively. These widths are optimized to encompass $\sim3.5$ standard deviations from the trace center. With the 1D stellar spectra, we then trim a handful of pixels at the edges of NRS2 and the red edge of NRS1. For NRS1, we trim the first (blue-most) $575$ pixels, as the stellar spectrum of this M-dwarf does not extend out this far into the blue-end of the spectrum. Thus, including these pixels would just mean we are adding excess noise to the spectrum. We also trim the first 100 integrations from all observations.

For the white lightcurve fitting, we first sum the 1D stellar spectra in the x-direction to get the white light flux (in counts per second) for each integration. Initially, we fit the white lightcurve separately for each detector and observation (eight total fits: four observations, with NRS1 and NRS2 fitted separately in each case). We use \texttt{batman} \citep{kreidberg2015batman} to fit the transit, fitting for the transit depth $(R_P/R_{\star})^2$, $a/R_{\star}$, impact parameter $b$, and mid-transit time $T_0$, while setting $e=0$ and $P=2.2531136$ days. At first, we also attempted to fit for the quadratic limb darkening coefficients, using the $q_1$ and $q_2$ parameterization put forth by \citep{kipping2013limbdark}, but these coefficients tended toward the same fixed values across all eight scenarios, so we instead fixed $q_1=0.1$ and $q_2=0$. At wavelengths as red as those in the G395H bandpass, limb darkening is quite constant and minimal, as evidenced by Fig.~\ref{fig:ld_comparison}.

We also need a systematics model to properly fit the transit. The out-of-transit data is used to measure the instrument systematics. We test every possible combination of systematics and use the Bayesian Information Criterion (BIC) to determine the best-fit model for each observation and detector. As shown in Fig.~\ref{fig1} and in the Allan deviation plots in Appendix~\ref{app:allan}, there are meaningful low-amplitude systematics seen in the white lightcurves, meaning there are undulations in the white light flux that vary smoothly and slowly with time. The magnitude and extent of these undulations varies between visits and detectors. Unlike the \texttt{Eureka!} pipeline, which handles this by trimming the first 640 integrations, we attempt to account for the undulations by applying a complex systematics model independently to each visit/detector. We use up to a 6th order polynomial in time, as well as x-shift and y-shift in a couple of cases. If we did not include such a complex model, and instead only following the model followed by the BIC, we ended up with a high amount of correlated noise.  

The undulations may be due to the the thermal cycling of the electronics and/or the low number of groups per integration (three in this case). A low number of groups per integration may impact the precision of the up-the-ramp read, and thus affect the flux over time. Indeed, there appears to be an inverse correlation between the amplitude of time-correlated noise in JWST near-infrared lightcurves and the number of groups implemented in each integration \citep[e.g.][]{alderson2024,hu2024,wallack2024}. Regardless, the use of a complex systematics model does seem to account for the undulations we see in our data. However, there are downsides to using a complex systematics model. The first is that we had difficulty determining consistent $a/R_{\star}$, $b$, and $T_0$ values across all observations. There is an inherent degeneracy between $a/R_{\star}$ and $b$, and it seemed that the different fits kept finding different regions of the degenerate parameter space of solutions. Typically, once the \texttt{FIREFLy} white lightcurves are fit individually using \texttt{emcee} \citep{foremanmackey2013emcee}, we take the weighted mean of $a/R_{\star}$ and $b$ and fix the white lightcurves to these weighted values before refitting for $(R_P/R_{\star})^2$ and $T_0$. However, in this instance, we had to fix $T_0$ in order to converge on a single solution for $a/R_{\star}$ and $b$. We approach this in a piecemeal manner: we first fix $T_0$ to its weighted mean, then refit the white lightcurves. Next, we fix $a/R_{\star}$ to its new weighted mean from the updated fits, refit the white lightcurves, and finally fix $b$ to its updated weighted mean. 

Finally, we fit the spectroscopic lightcurves. \texttt{FIREFLy} employs a binning scheme based on (roughly) equal counts per bin, instead of equal wavelength spacing. We also produced an additional \texttt{FIREFLy} reduction with the same $\Delta \lambda = 0.04~\mu$m-binning scheme as \texttt{Eureka!} in order to directly compare both reductions in Fig.~\ref{transmission_spectra_comparison_reductions} (all atmospheric retrievals on the \texttt{FIREFLy} data in Section~\ref{sec:retrievals} used the equal-counts-per-bin reduction). We bin the 1D stellar spectra according to the binning scheme, then plot the relative change in flux with time per spectroscopic bin. We then fit these spectroscopic lightcurves by fixing $a/R_{\star}$, $b$, $T_0$, and limb darkening coefficients to their white lightcurve values. We also fix the systematics model to the white lightcurve values, as the systematics are quite gray (wavelength-independent). The one exception to this is the linear term in the polynomial, which does have a wavelength dependence and is fit for in the spectroscopic lightcurves.

The final step we had to take was applying offsets to some of the resulting spectra in order for NRS1 and NRS2 to align. Detector offsets in NIRSpec G395H are possible and have been investigated \citep{madhu2023k218b, may2023gj1132b}, and some of the offsets we see are much too large to be physical - on the order of 50--100 ppm. This is likely driven by our complex systematics model, which can obfuscate the true transit depth and contribute to the differences in the retrieved white light-curve values of $R_p/R_s$ in Table~\ref{tab2}. Because transmission spectra are only concerned with relative differences, this is not an issue when examining one spectra, but when stacking multiple spectra (say, across detectors and visits), this can lead to arbitrarily different mean depths. In order to combat this, then, we find that applying offsets so that the mean transit depth across all detectors and visits is consistent. We elect to use Visit 2, where NRS1 and NRS2 agree quite well, with a mean depth of 639 ppm for the entire visit, and offset all other spectra to this mean value. We note that \texttt{Eureka!} does not employ this technique, and yet our spectra still agree quite well (Fig.~\ref{transmission_spectra_comparison_reductions}).

\section{Comparison to forward models}
\begin{figure*}
\centering
\includegraphics[width=\textwidth]{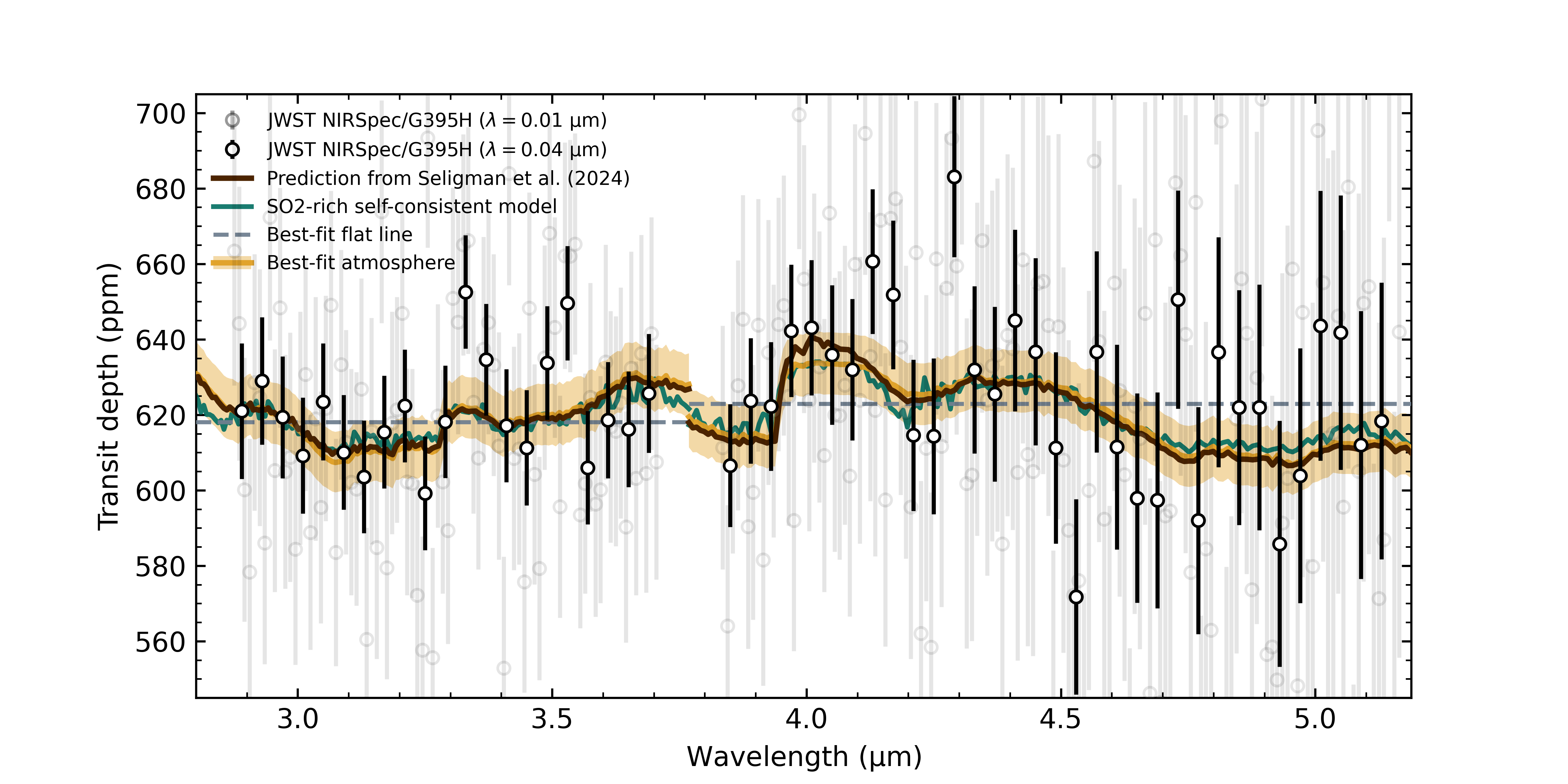}
\caption{The average \texttt{Eureka!} transmission spectrum of \planetname\ from the four JWST NIRSpec/G395H transits compared against the 98\% \ce{SO2} model predicted in \citet{seligman2024tidal} and a self-consistent photochemical model assuming an \ce{SO2}-dominated atmosphere. We also show the best-fit flat line and atmosphere models retrieved with Aurora on the $\Delta \lambda = 0.01~\mu m$ data and the corresponding 2$\sigma$ uncertainty bands. All models include an offset between the two detectors.}\label{fig2}
\end{figure*}

To interpret the observations, we compare the weighted average of the four \texttt{Eureka!} transmission spectra (shown in Fig.~\ref{fig2})
against a series of models, ranging from a bare rock to self-consistent \ce{SO2}-dominated atmospheric models.

\subsection{Airless model}
We begin with the simplest possible model of a bare rock with no atmosphere, that is, a flat line with two free parameters: a constant transit depth and an offset between the two detectors. The flat line, shown in Fig.~\ref{fig2}, provides a reasonable fit to the data with a $\chi^2=197.84$ (216 degrees of freedom) for $\Delta\lambda = 0.01~\mu$m (218 data points).

\subsection{Volcanic atmospheres from \citet{seligman2024tidal}}
We then compute the $\chi^2$ value relative to the three synthetic spectra of volcanic atmospheres in \planetname\ presented in \citet{seligman2024tidal}, which are predominantly composed of varying amounts of \ce{SO2} and \ce{CO2}. We allow for two free parameters: a general vertical offset to account for differences in the assumed reference pressure in the prediction models, and an offset between the two detectors. The corresponding $\chi^2$ values are 203.9, 191.6, and 189.0 for the 5\%, 50\%, and 98\% \ce{SO2} synthetic spectra, respectively (216 degrees of freedom). In Fig.~\ref{fig2}, we compare the JWST transmission spectrum of \planetname\ against the 98\% \ce{SO2} model from \citet{seligman2024tidal}.

\subsection{Self-consistent forward model from EPACRIS}\label{subsec:models}
We further assess the goodness of fit with a self-consistent photochemical model of an \ce{SO2}-dominated atmosphere under radiative-convective equilibrium using the ExoPlanet Atmospheric Chemistry \& Radiative Interaction Simulator (EPACRIS, Scheucher et al., in preparation, see Appendix~\ref{app:epacris} for more details). We allow for the same two vertical offsets as in the previous case. The resulting transmission spectrum results in a $\chi^2=188.9$ (216 degrees of freedom), also providing a good explanation for the data (Fig.~\ref{fig2}).

\begin{figure*}
\centering
\includegraphics[width=\textwidth]{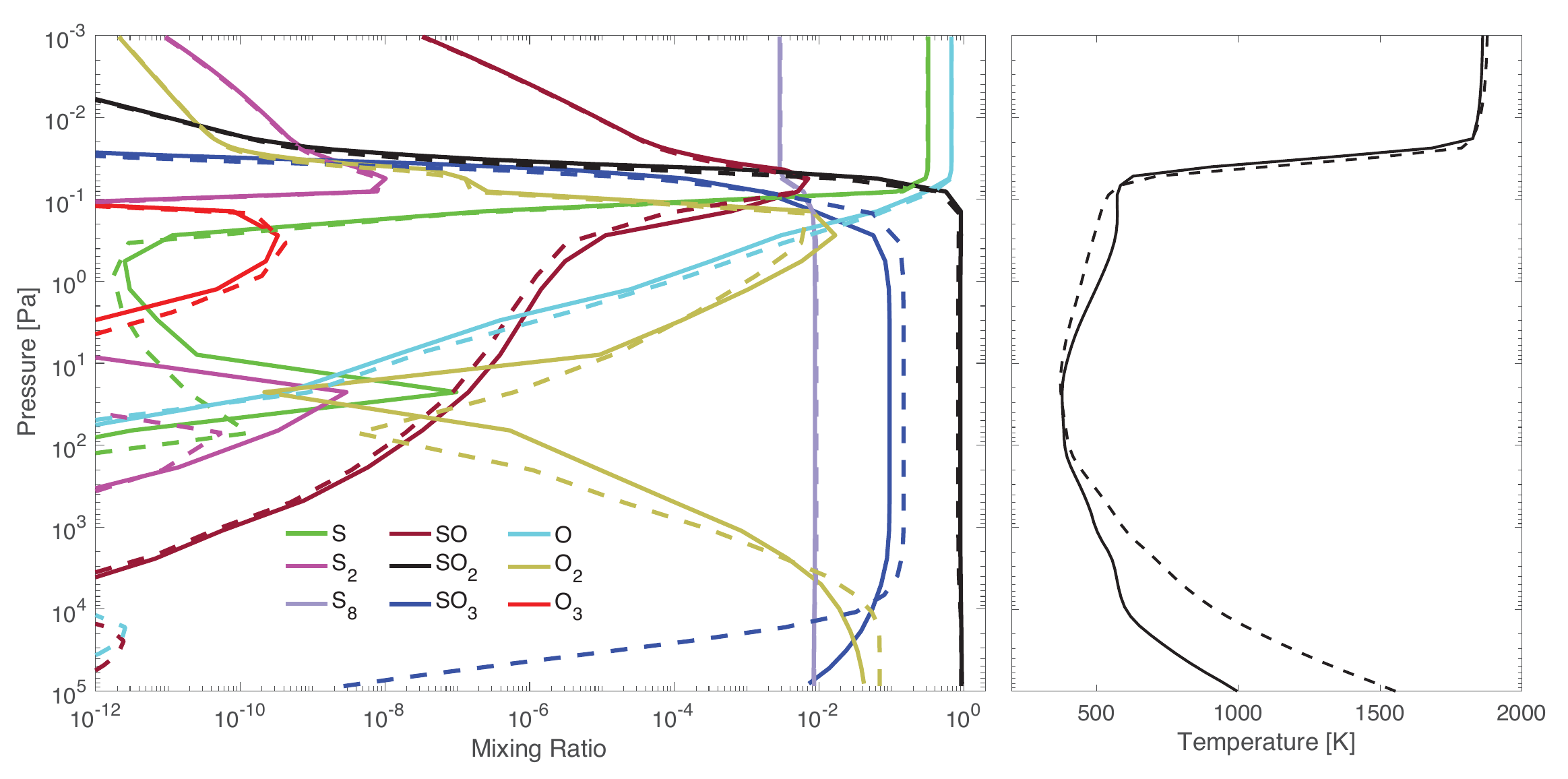}
\caption{Self-consistent models of an \ce{SO2}-dominated atmosphere on \planetname. The figure shows the mixing ratio of key molecules (left) and the temperature (right) as a function of pressure, and the solid and dashed lines correspond to an internal heat flux of $1\times$ and $10\times$ the insolation, respectively, corresponding to a tidal Q value of 30 and 3. The atmosphere should build up abundant \ce{SO3} and gas-phase elemental sulfur (\ce{S8}) produced from the photolysis of \ce{SO2}.}
\label{fig_forward}
\end{figure*}

As shown in Fig.~\ref{fig_forward}, our self-consistent models indicate that an \ce{SO2}-dominated atmosphere on \planetname\ will have concurrent high abundances of \ce{SO3} and elemental sulfur (represented as \ce{S8} in our models).
The visible-wavelength absorption of elemental sulfur causes a moderate temperature inversion at $~10$ Pa (Fig.~\ref{fig_forward}), while the entire middle atmosphere ($10^{-1}\sim10^{3}$ Pa) would have a temperature of approximately 400 K, which is substantially lower than the planet's zero-albedo equilibrium temperature and little impacted by the assumed tidal heating rate.

\section{Atmospheric retrievals}\label{sec:retrievals}
To further assess the detection significance, we performed Bayesian atmospheric inferences with a series of models and tools: \texttt{ExoTR} \citep{damiano2024lhs}, Aurora \citep{Welbanks2021}, and POSEIDON \citep{MacDonald2017, MacDonald2023}.

\subsection{ExoTR}\label{subsubsec:exotr}
\texttt{ExoTR} \citep[Exoplanetary Transmission Retrieval, ][]{damiano2024lhs} is a fully Bayesian retrieval algorithm designed to interpret exoplanet transmission spectra. Some of the capabilities of \texttt{ExoTR} include: a) the cloud layer can be modeled as an optically thick surface or as a physically motivated cloud scenario tied to a non-uniform water volume mixing ratio profile, similarly to \texttt{ExoReL$^\Re$} \citep{hu2019information, damiano2020exorel, damiano2022small}, b) the stellar heterogeneity components can be jointly fit with the planetary atmospheric parameters \citep{rackham2017heterogeneity, pinhas2018retrieval}, c) the atmospheric abundances are fit in the centered-log-ratio (CLR) space and the prior functions are designed to render a flat prior when transformed back to the log-mixing-ratio space \citep{damiano2021prior}, and d) the possibility to fit photochemical hazes with prescribed optical constants and a free particle size. \texttt{ExoTR} will be described in detail in a subsequent paper (Tokadjian et al. in prep.). 

Table~\ref{tab:exotr_setup} in Appendix~\ref{app:exotr} lists the free parameters, the prior space used, and the range in which the parameters are probed. We defined the offsets relative to the datasets as the $\Delta$ppm relative to the G395H NRS1 dataset (chosen as reference). The planetary temperature is modeled as an isothermal and the clouds as an optically thick layer. The stellar heterogeneity has been modeled by following the prescription presented in \citet{pinhas2018retrieval}. The concentrations of the gases have been explored in the centered-log-ratio (CLR) space, which allows for any trace gas to become the background gas and is thus the most agnostic in terms of bulk atmospheric composition \citep{Benneke2012}. These concentrations have been converted into volume mixing ratio (VMR) in the results presented here. \texttt{ExoTR} uses \texttt{MultiNest} \citep{feroz2009multinest} to sample the Bayesian evidence, estimate the parameters, and determine the posterior distribution functions. \texttt{MultiNest} is used through its \texttt{Python} implementation \texttt{pymultinest} \citep{buchner2014multinest}. For all the retrieval analyses presented here, we used 800 live points and 0.5 as the Bayesian evidence tolerance. Finally, to assess the significance of a scenario over the null hypothesis, we calculated the Bayes factor \citep{trotta2008bayes}, which is a quantitative statistical measurement to choose one model over another one.

\begin{figure*}
\centering
\includegraphics[width=\textwidth]{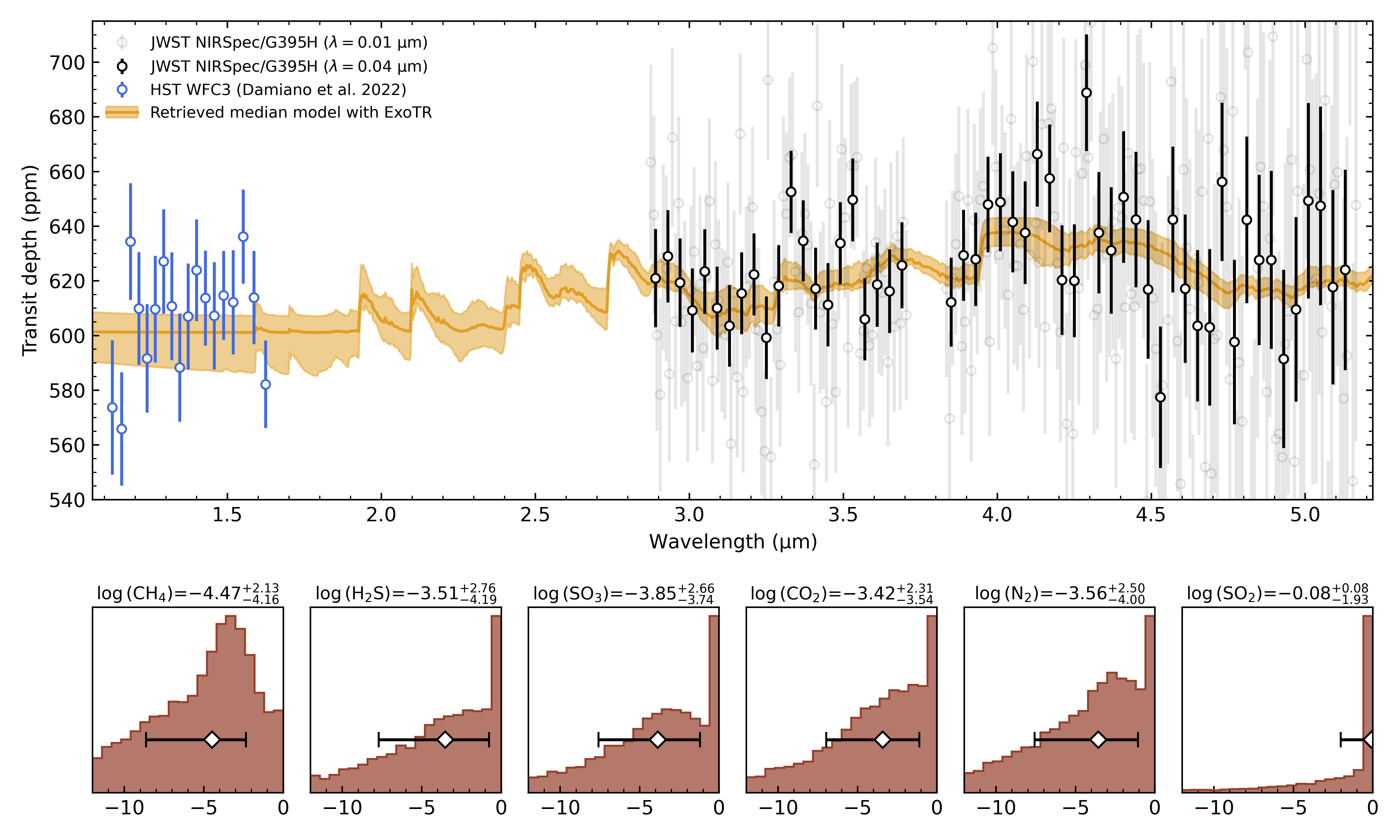}
\caption{ExoTR retrieval results for \planetname. \textit{Top}: retrieved mean spectrum (mean model and 2$\sigma$ confidence interval) from scenario 8 in Table~\ref{tab:exotr_res}. \textit{Bottom}: Posterior distribution functions for selected gases. The posterior distribution functions suggest a heavy atmosphere rich in \ce{SO2}. \label{fig:exotr_retr}}
\end{figure*}

In the analysis using \texttt{ExoTR}, we used the \planetname\ transmission spectrum from \texttt{Eureka!} at $\Delta \lambda = $ 0.01. We started our analysis by running \texttt{ExoTR} on the combined dataset (HST$+$JWST). The initial step was to define a baseline that would serve as null-hypothesis, for this reason, we run a bare rock scenario (see Table \ref{tab:exotr_res} in Appendix, scenario 11) which only has offsets between the datasets and the planetary radius and no atmospheric parameters. With the baseline scenario defined, we then run multiple scenarios in which we included multiple gases (i.e., H$_2$O, CH$_4$, H$_2$S, CO$_2$, \ce{SO2}, \ce{SO3}, and N$_2$ as filler gas), clouds, and stellar heterogeneity (see Table \ref{tab:exotr_res}, scenarios 8, 9, and 10). We find that Scenario 8, which includes seven different gases, clouds, and temperature as free parameters, is preferred over the flat-line model by 3.37$\sigma$. It is also worth noting the following trends: (i) clouds and stellar heterogeneity do not contribute to significantly enhance the interpretation of the spectrum, (ii) absorption features from \ce{SO2} could be present, moreover, it seems that \ce{SO2} might be identified as the dominant gas of the atmosphere (see Figure \ref{fig:exotr_retr}).
When \ce{SO2} is removed from the set of free parameters, the Bayesian evidence of the fitted model is reduced. Because of the significance of \ce{SO2} and predictions from self-consistent models, we also run two more retrievals with the addition of SO$_3$ as a fitting gas (scenarios 2 and 3 in Table \ref{tab:exotr_res}). Even if the self consistent calculations suggest the presence of SO$_3$, there is not a significant increase in the evidence when adding it as free parameter.

With these findings, we then moved to analyze the JWST data only. Also in this case, we defined the baseline scenario, i.e., bare rock (scenario 6 in Table \ref{tab:exotr_res}) by only defining the offset between the NR1 and NRS2 of the G395H dataset, and the planetary radius as free parameters. We did not include stellar heterogeneity in our analysis of JWST data only. We initially explored a retrieval that included all the gas listed in Table \ref{tab:exotr_setup} as free parameters. Also in this case, H$_2$O and CH$_4$ are unconstrained and we decided to drop them from subsequent trials. We noticed that if the planetary temperature is considered as a free parameter, lower values ($\sim$250 K) are preferred compared to the equilibrium temperature. We then proceed to exclude all the gases but \ce{SO2} as it appeared to be the most likely absorber in the atmosphere, and we found that if we compare the baseline case (scenario 6) with a scenario that has 100\% \ce{SO2} and has the same number of free parameters as the bare rock model, i.e., offset and planetary radius, we calculated a significance of 3.53$\sigma$ (see Table \ref{tab:exotr_res}, scenario 1). Similar to the HST+JWST case, we include scenarios with the addition of SO$_3$, and we do not observe a substantial change in the evidence (scenario 8). With the addition of NH$_3$ and CO, we do not gain any appreciable evidence. Once again, adding clouds does not impact the retrieval result. The results from scenario 1 and 2 translates into a moderate preference of \ce{SO2} absorption in the atmosphere of \planetname. This result should not be considered as a detection, indeed, the presence of \ce{SO2}, even though preferred with higher likelihood, is degenerate with the presence of N$_2$ which instead would result in a flat model. From this analysis, we could not assign any significant evidence to other gases. Supplemental observations are needed to enhance the significance to over 5$\sigma$ when comparing an atmospheric model to a flat line \citep{seligman2024tidal}.

\subsection{Aurora, Cross-Validation, and Self-Consistent Model Analysis}\label{subsubsec:aurora_loocv}

We perform an additional set of Bayesian inferences using Aurora \citep{Welbanks2021}, a framework developed for the analysis of transmission \citep{Welbanks2024} and emission spectroscopy \citep{Bell2023}. We focus our analysis on the \texttt{Eureka!} reduction. To interpret the transmission spectra of \planetname, Aurora solved radiative transfer for a parallel-plane atmosphere in transmission geometry assuming hydrostatic equilibrium. The vertical temperature structure of the planet considered ranged from a simple isothermal treatment to the parametric treatment from \citet{Madhusudhan2009}. We consider the presence of inhomogeneous clouds and hazes as a linear combination of cloudy, hazy, and clear models as outlined in \citet{Welbanks2021}. Similarly, with Aurora we account for the impact of stellar contamination in the observed spectrum following the implementation outlined \citet{pinhas2018retrieval}. The sources of opacity considered are obtained from HITRAN \citep{Rothman2010, Richard2012} and ExoMol \citep{Tennyson2016} as described in \citet{Welbanks2021, Welbanks2024}. Finally, Aurora is a generalized retrieval framework that can relax the assumption of an H-rich atmosphere in the analysis of any spectra. This is done by implementing tools from compositional data analysis such as the use of the centered-log-ratio transformation for the priors on the molecular abundances in the atmosphere. The parameter estimation is performed using nested sampling through MultiNest \citep{buchner2014multinest, feroz2009multinest}.

We begin by interpreting the observations using a flat-line model. We use a model with two free parameters, a transit depth with a uniform prior between 0.02\% and 0.09\%  and an offset between the NRS1 and NRS2 observations with a Gaussian prior centered at zero with a standard deviation of 100 ppm. On the combined \texttt{Eureka!} observations from the four transits at $\Delta\lambda = 0.01~\mu$m, this model results in a minimum $\chi^2=197.84$. Assuming 216 degrees of freedom, the resulting p-value ($p=0.81$) cannot rule out the null hypothesis and this flat line model is consistent with the observations.

Given the goodness-of-fit of the flat line model, a more complex model (e.g., with more free parameters) can over-fit the data. An atmospheric retrieval considering a model with stellar activity, fully inhomogeneous clouds and hazes, and non-isothermal vertical temperature structure, would have almost as many free parameters as there are spectral bins in our observations. Therefore, we perform a subsequent analysis with an intermediate 14 parameter model: 8 gas species (i.e., H$_2$+He in solar proportions, H$_2$O, CH$_4$, CO, CO$_2$, H$_2$S, N$_2$, and  \ce{SO2}), an isothermal temperature, one parameter for the pressure of an optically thick cloud deck, one for a cloud cover fraction, one for the reference pressure, one for the planetary radius at the reference pressure, and one for an offset for NRS1 relative to NRS2. The atmospheric models for Aurora are computed at a resolution of 20,000 between 2.5~$\mu$m and 5.3~$\mu$m. 

\begin{figure*}
\centering
\includegraphics[width=0.9\textwidth]{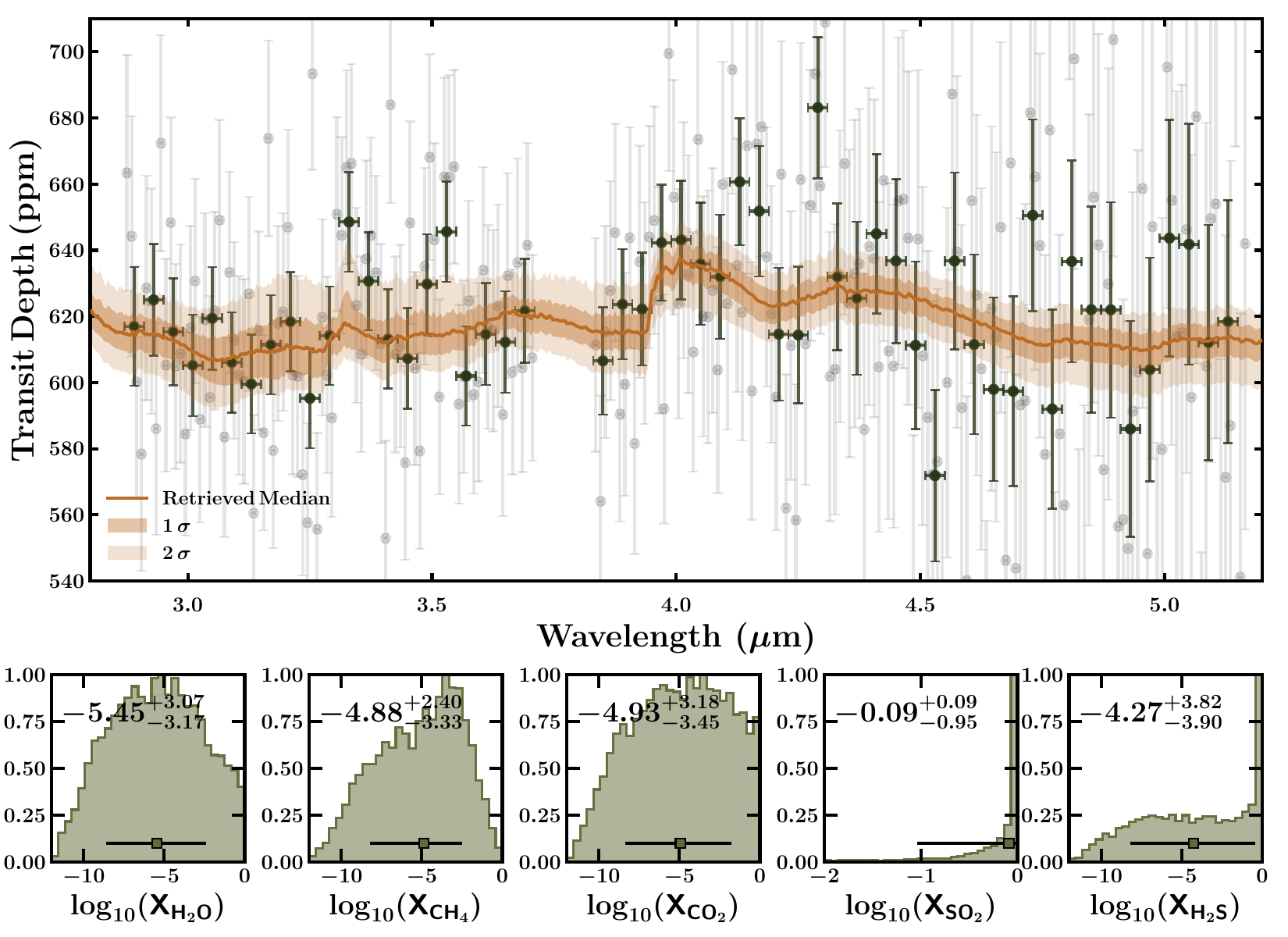}
\caption{Aurora retrieval results for \planetname. Top: The retrieved transmission spectrum on the Eureka! reduction of the \planetname\ observations. The inference is performed on the $\Delta\lambda = 0.01~\mu$m resolution observations, but the $\Delta\lambda = 0.04~\mu$m data overplotted for visual clarity. The orange shading shows the retrieved median model as well as 1$\sigma$ and $2\sigma$ confidence intervals. Bottom: Posterior distributions for the gases of interest. Most gases are unconstrained with inferred abundances suggestive of an atmosphere rich in SO$_2$ }
\label{fig:aurora_1_results}
\end{figure*}

The 14 parameter retrieval results in a $\chi^2=187.32$. The associated p-value under the assumption of 204 degrees of freedom, $p=0.79$, suggests that this atmospheric model is an appropriate fit to the data. Performing a Bayesian model comparison of this atmospheric model to the flat line model described above, we obtain a 3.3$\sigma$ model preference for the atmospheric model over the flat line model. We clarify that this comparison is based solely on the Bayesian evidence of the models. The atmospheric retrieval does not place meaningful constraints on the chemical abundances of the gases, vertical temperature structure of the planet, or cloud/haze properties. The only absorber preferred by this atmospheric model is SO$_2$ at the 2.4$\sigma$ level based on the comparison of this reference model to a nested model without SO$_2$. Further comparisons with a cloud-free model and models considering stellar activity result in no meaningful preference for these effects. Figure \ref{fig:aurora_1_results} shows the retrieved transmission spectra and retrieved abundances for the gases of interest. While unconstrained, the retrieval allows for large abundances of \ce{SO2}, making this absorber the main constituent of the atmosphere. The retrieved isothermal temperature, $\text{T}_\text{iso}=596^{+125}_{-143}$~K, is consistent with the equilibrium temperature of the planet. In Fig.~\ref{fig:ratio_so2_co2}, we present the retrieved values of the \ce{SO2}/\ce{CO2} abundance ratio to facilitate comparison with the different atmospheric scenarios presented in \citet{seligman2024tidal}.

As with the ExoTR analysis above, we consider the possibility of an atmospheric model with 100\% \ce{SO2} at the equilibrium temperature of the planet, with two free parameters: one for the planetary radius and one for an offset between NRS1 and NRS2. The parameter for the planetary radius is set at a reference pressure of 1 bar and acts as a vertical offset for the spectrum \citep{Welbanks2019a}. This two parameter model aims to serve as a comparison to the flat-line model fit above given the same number of degrees of freedom. This simple retrieval results in a $\chi^2=189.13$ (216 degrees of freedom), smaller than that of the flat-line model and still not rejected by the p-value ($p=0.91$). A Bayesian model comparison between this simple atmospheric model and the flat-line model above results in a 3.6$\sigma$ model preference, based on the Bayesian evidence, for the atmospheric model over the flat line model. Even though the retrieved interdetector offsets are small and consistent with zero to within 1$\sigma$ (e.g. $\textup{NRS1}-\textup{NRS2}=-4.8^{+5.4}_{-5.1}$ and $4.5^{+10.6}_{-6.8}$ ppm for the flat-line and pure \ce{SO2} atmosphere models, respectively), we ran an additional set of retrievals in which we did not allow for an offset between detectors. In this case, the preference for the \ce{SO2} atmospheric model over the flat-line model in terms of Bayesian evidence increases to 3.8$\sigma$.

Given the signal-to-noise of the observations, and based on their resulting p-values, we cannot definitely reject the flat-line model, the 14-parameter atmospheric model, or the simple two-parameter atmospheric model. We perform the same set of retrievals and model comparisons on the $\Delta\lambda = 0.02~\mu$m and $\Delta\lambda = 0.04~\mu$m data and find that our inferences are consistent across all three data resolutions. The analysis of the model evidence suggests that an atmospheric model is preferred over the flat line, with the highest model preference corresponding to the simple model over the flat line at $3.6\sigma$. However, these model assessments metrics provide a single value shedding little light on which points drive these specific preferences \citep{Welbanks2023}.

\begin{figure}
\centering
\includegraphics[width=0.5\textwidth]{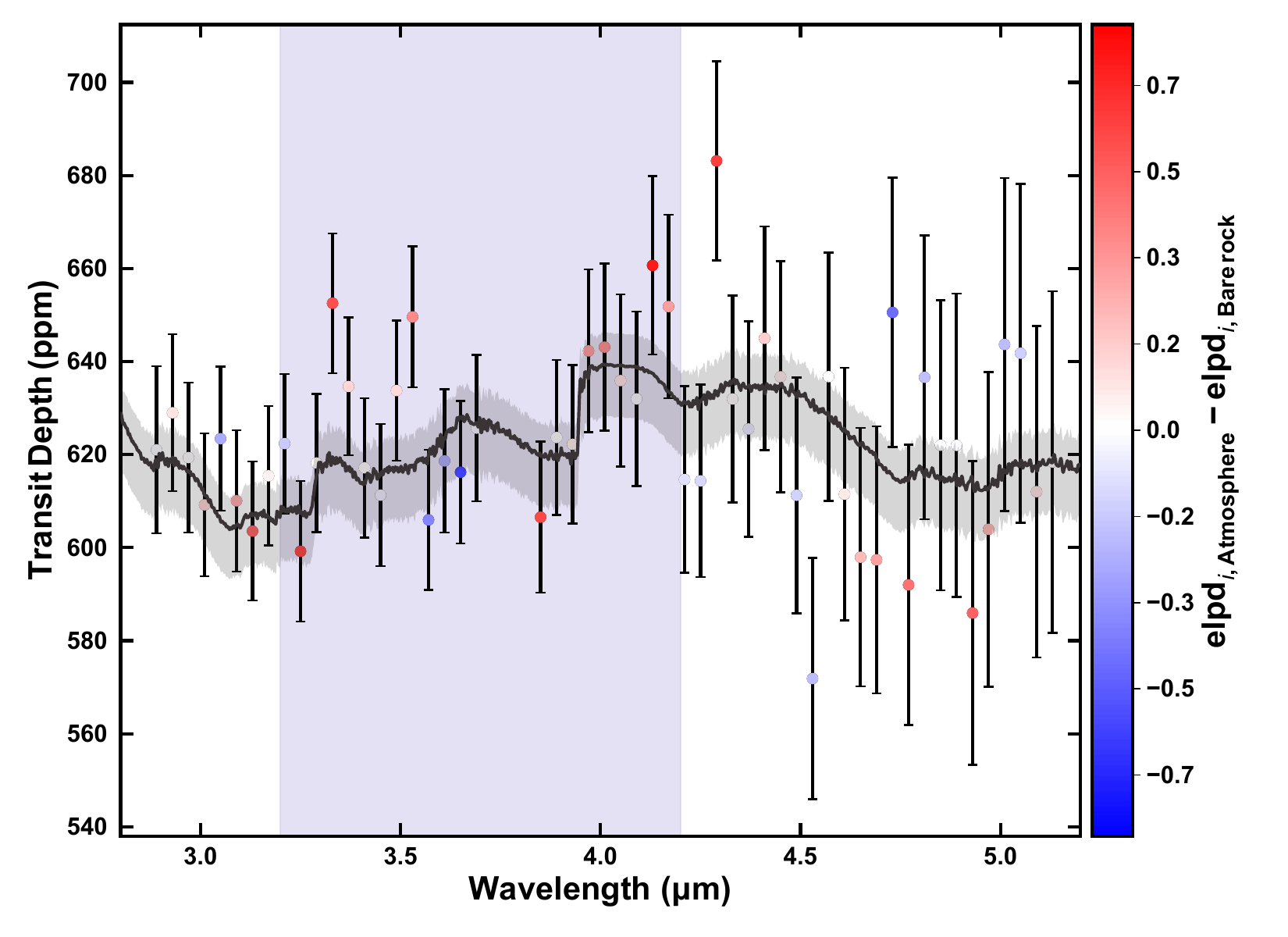}
\caption{The model preference for a planet with an atmosphere over a flat line is consistent with regions of dominant SO$_2$ absorption. The data is color coded by the point-wise difference in the expected log point-wise predictive density (elpd) between the simple atmospheric model and the bare-rock (that is, flat line) model. The retrieved median and 2$\sigma$ confidence interval from the simple atmospheric model are shown in gray. Redder data points, that is those with larger positive $\Delta$elpd, are better explained by the atmospheric model with SO$_2$ absorption.}
\label{fig:LOO_scores}
\end{figure}

We study the impact of offsets in the overall $R_p/R_s$ derived per visit by performing retrievals on a modified combined spectrum. Namely, instead of taking the weighted average of the four individual spectra, we first subtract from each spectrum the $R_p/R_s$ value derived from the corresponding white lightcurve fit. Then, we add the mean $R_p/R_s$ value from the four white lightcurves of the corresponding detector. The conclusions remain largely unchanged: a pure \ce{SO2} atmosphere is still preferred over a flat line based on its Bayesian evidence (3.4$\sigma$), but we cannot definitely rule out a no-atmosphere model, that is a flat line. In the case of the 14-parameter retrieval, the 1$\sigma$ lower limit on $\log(\ce{SO2})$ abundance decreases to -6.4 instead of -1.0.

To further compare these atmospheric models to the flat line model, we turn to Bayesian Leave-One-Out Cross-Validation \citep{McGill2023, Welbanks2023}. We perform a per datum comparison between the simple atmospheric model and flat line model to determine which regions in the spectrum drive this model preference. Generally, in LOO-CV a model is trained on the dataset leaving out one data point at a time, and scoring how well the trained model can predict the left out data point (that is, the expected log predicted density of the left out datum - elpd). The process is performed for all data points in the spectrum and the out-of-sample predictive performance of the model is estimated. Fig.~\ref{fig:LOO_scores} shows the difference in elpd scores between the simple atmospheric model and the bare rock (that is, flat-line) model. The difference between the scores shows where one model outperforms another. 

The LOO-CV analysis shown in Fig.~\ref{fig:LOO_scores} was performed on the $\Delta\lambda = 0.04~\mu$m data. Of the top five points with the highest scores, four are in regions where SO$_2$ is the main absorber when compared to CO$_2$. The only exception is the point with the second highest score, at $\sim4.13~\mu$m. Our LOO-CV analysis finds that the density of the increased predictive performance \citep[that is, sum($\Delta$elpd)/\#points,][]{Welbanks2024} is higher and over double the value in regions where SO$_2$ is the dominant cross-section relative to CO$_2$, than in regions where SO$_2$ is not dominant. The atmospheric model results in an increase in the predictive performance at 2.2 standard errors over the bare rock model.  

We perform a final analysis to assess the goodness of fit using the self-consistent forward models with EPACRIS coupled with Aurora. The description of the self-consistent model is presented in Section~\ref{subsec:models} and Appendix~\ref{app:epacris}. For this exercise, we perform retrievals using the radiative-convective equilibrium vertical temperature structure from the 1$\times$ and 10$\times$ insolation models with a $\Delta T$ parameter to allow for deviations from this equilibrium profile. While the self-consistent models result in an SO$_2$ rich atmosphere, we allow for the abundance of SO$_2$ to be a free parameter with log-uniform priors between -12 and 0. We allow for the rest of the atmosphere to be filled with CO$_2$ gas to assess whether the data prefers larger abundances than that produced by the self-consistent models. We include one free parameter for the planetary radius at 10 bar, and a free parameter for the offset between NRS1 and NRS2. This exercise is performed on the $\Delta\lambda = 0.01~\mu$m data.

\begin{figure}
\centering
\includegraphics[width=0.4\textwidth]{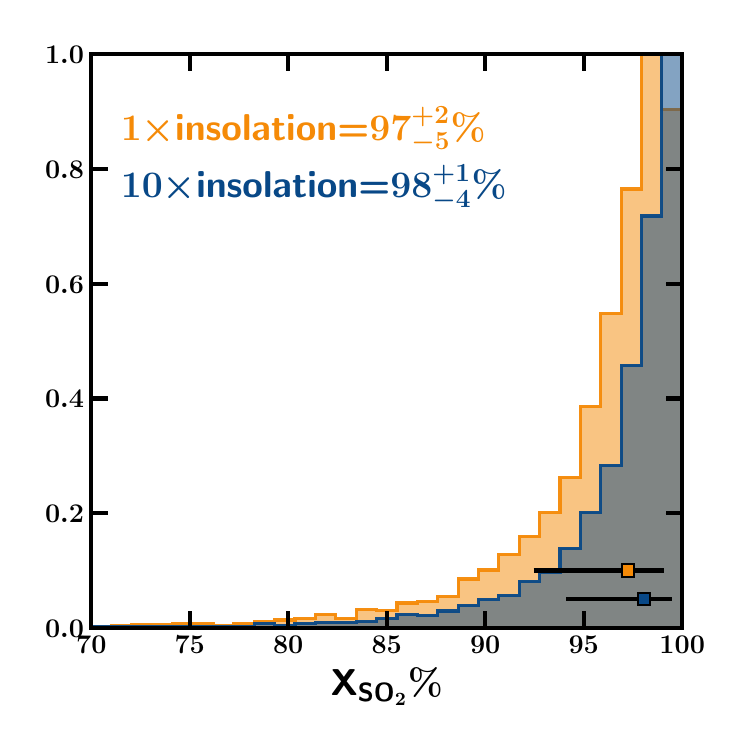}
\caption{Retrieved SO$_2$ abundances for models using radiative-convective equilibrium vertical temperature structures. The $1\times$ and $10\times$ insolation models infer SO$_2$ abundances over 90\% within their 68\% confidence interval, favoring a SO$_2$-rich atmosphere and disfavoring large CO$_2$ abundances.  }
\label{fig:so2_abundance}
\end{figure}

The atmospheric retrievals result in a $\chi^2=188.04$ and $\chi^2=187.88$ for the $1\times$ and $10\times$ insolation models, respectively. The inferred $\Delta T$ are largely unconstrained and consistent with no variation, that is  $\Delta T=0$~K, to the radiative-convective vertical temperature structure from EPACRIS. The atmospheric retrievals find a strong constraint on the SO$_2$ abundance of $97^{+2}_{-5}\%$ and $98^{+1}_{-4}\%$ for the $1\times$ and $10\times$ models, respectively. A Bayesian model comparison suggests a weak preference at $\sim2\sigma$ for the 10$\times$ insolation vertical temperature structure model over that with the 1$\times$ insolation vertical temperature structure. Fig.~\ref{fig:so2_abundance} shows the retrieved posterior probability distributions for the retrievals using the self-consistent models' vertical temperature structure.

\subsection{POSEIDON}\label{subsubsec:poseidon}

We conducted an additional retrieval analysis of \planetname's transmission spectrum with the open source retrieval code POSEIDON \citep{MacDonald2017, MacDonald2023}. We consider three retrieval scenarios: (i) a flat line, (ii) stellar contamination from unocculted stellar inhomogeneities, and (iii) a planetary atmosphere. All three scenarios allow for a relative offset between the NIRSpec G395H NRS1 and NRS2 detectors. We initially performed retrievals with POSEIDON on the individual visit transmission spectra, but we found no evidence of any spectral deviations from a flat line. Therefore, in what follows, we present results for the combined 4-visit transmission spectrum of \planetname. We repeated our POSEIDON retrieval analysis for both the \texttt{Eureka!} and \texttt{FIREFLy} data reductions and for the three \texttt{Eureka!} data bin sizes ($\Delta \lambda = 0.01~\mu$m, $0.02~\mu$m, and $0.04~\mu$m).

Our three POSEIDON retrieval models span a range of astrophysical scenarios to explain \planetname's transmission spectrum. First, the flat line model corresponds to a rocky body with no appreciable atmosphere transiting a star with negligible stellar activity. The 2-parameter flat line model is defined by the planetary radius ($R_p \sim \mathcal{U}$(0.85\,$R_{\rm{p, \, obs}}$, 1.15\,$R_{\rm{p, \, obs}}$, where the observed radius is 0.85\,$R_\earth$) and an instrumental systematic offset between the NRS1 and NRS2 detectors ($\delta_{\rm{rel}} \sim \mathcal{U}$(-200\,ppm, +200\,ppm)). The stellar contamination model similarly assumes an atmosphere-less planet but accounts for unocculted stellar spots and faculae outside the transit chord. The 7-parameter stellar contamination model adds five additional free parameters (priors in brackets): the stellar photosphere temperature ($T_{\rm{phot}} \sim \mathcal{N}$($T_{*, \rm{eff}}$, $\sigma_{T_{*, \rm{eff}}}$), the spot/faculae covering fractions ($f_{\rm{spot}}$/$f_{\rm{fac}} \sim \mathcal{U}$(0.0, 0.5)), and the spot/faculae temperatures ($T_{\rm{spot}} \sim \mathcal{U}$(2300\,K, $T_{\star,\,\rm eff} + 3\,\sigma_{\rm T_{\star,\,\rm eff}})$, $T_{\rm{fac}} \sim \mathcal{U}(T_{\star,\,\rm eff} - 3\,\sigma_{\rm T_{\star,\,\rm eff}}, 1.2\,T_{\star,\,\rm eff})$). For the stellar parameter priors, we adopt literature properties of the host star: $T_{*, \rm{eff}} = 3412$\,K and $\sigma_{T_{*, \rm{eff}}} = 49$\,K \citep{cloutier2019l9859}. We calculate stellar contamination spectra by interpolating PHOENIX model spectra \citep{Husser2013} via the \texttt{PyMSG} package \citep{Townsend2023}. Finally, we fit a model with a planetary atmosphere on \planetname\ without stellar contamination. The 11-parameter atmosphere model is defined by the atmospheric temperature ($T \sim \mathcal{U}$(100\,K, 800\,K)), the planetary mass \citep[$M_{p} \sim \mathcal{N}$(0.47\,$M_{\earth}$, 0.15\,$M_{\earth}$,][]{rajpaul2024}, the radius at the 1\,bar pressure level ($R_{\mathrm{p, \, ref}} \sim \mathcal{U}$(0.85\,$R_{\rm{p, \, obs}}$, 1.15\,$R_{\rm{p, \, obs}}$), the surface pressure, which mimics the physics properties of an opaque cloud deck ($\log_{10} (P_{\rm{surf}}\,/\,\rm{bar}) \sim \mathcal{U}$(-7, 2))), the NRS1--NRS2 free offset ($\delta_{\rm{rel}} \sim \mathcal{U}$(-200\,ppm, +200\,ppm)), and six free parameters encoding the volume mixing ratios of N$_2$, CO$_2$, SO$_2$, H$_2$O, CH$_4$, H$_2$S ($\log_{10} X_{i} \sim$ CLR(-12, 0), where `CLR' is the centred-log ratio prior, \citealt{Benneke2012}). The volume mixing ratio of H$_2$+He (with a fixed primordial ratio of He/H$_2$ = 0.17) fills any remaining primary atmosphere, also following a CLR prior, but is not a free parameter due to the summation to unity condition for mixing ratios. The molecular cross sections used by POSEIDON are computed from the following ExoMol \citep{Tennyson2016} line lists: CO$_2$ \citep{Tashkun2011}, SO$_2$ \citep{Underwood2016}, H$_2$O \citep{Polyansky2018}, CH$_4$ \citep{Yurchenko2017}, and H$_2$S \citep{Azzam2016}. We also consider collision-induced absorption (e.g. N$_2$-N$_2$ pairs) from HITRAN \citep{Karman2019}. We compute model spectra for all three models at a spectral resolution of $R = $ 20,000 from 2.6--5.3~$\mu$m and sample the parameter spaces with 2,000 MultiNest live points.

\begin{figure*}
\centering
\includegraphics[width=0.9\textwidth]{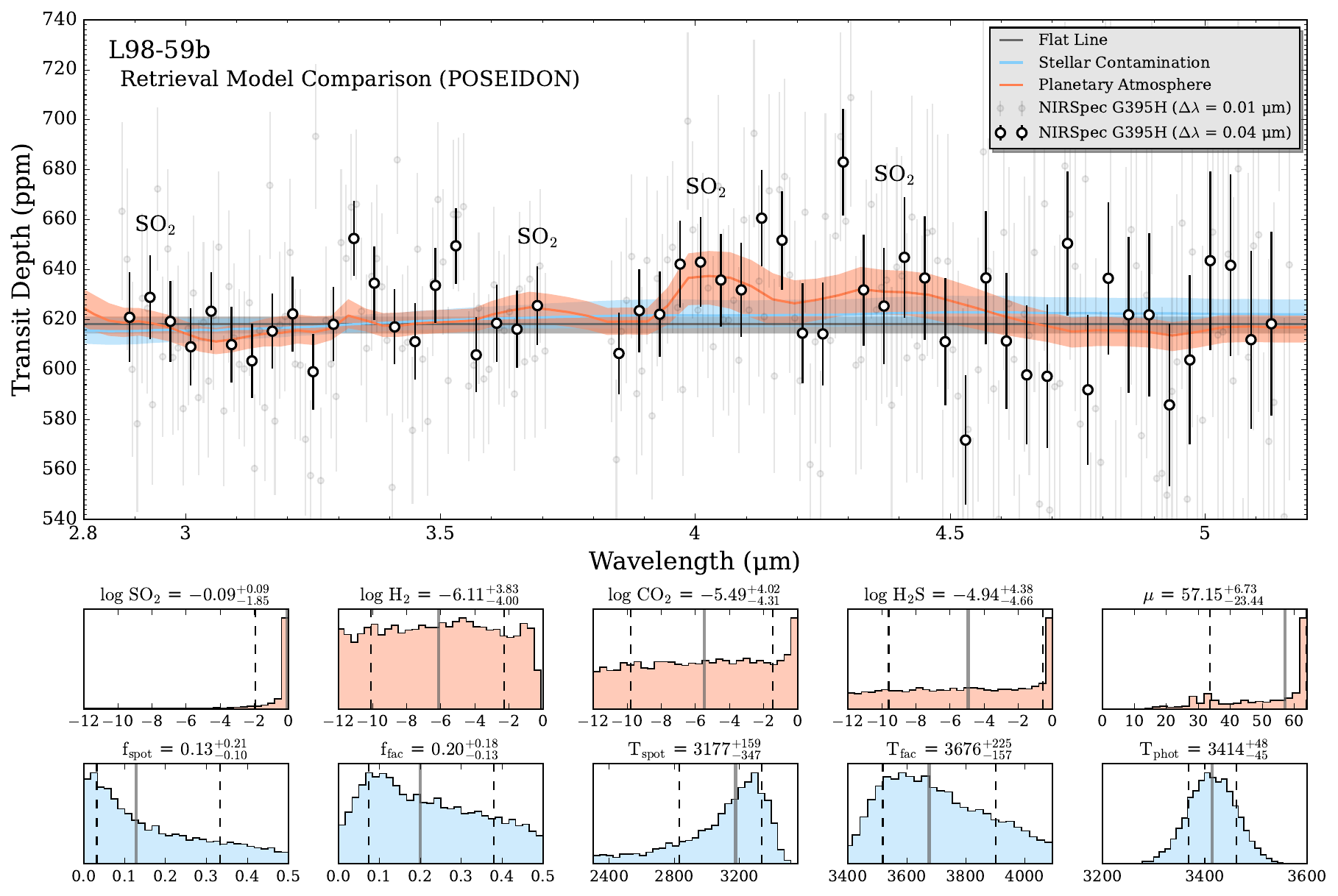}
\caption{POSEIDON retrieval results for \planetname. Top: retrieved transmission spectra (median model and 1$\sigma$ confidence interval) for three models: a flat line (grey), stellar contamination (blue), and an SO$_2$-rich planetary atmosphere (orange). The retrieval models shown correspond to fits on the highest-resolution ($\Delta \lambda = 0.01~\mu$m) \texttt{Eureka!} data (transparent error bars), with the lower-resolution $\Delta \lambda = 0.04~\mu$m data overplotted for clarity (non-transparent error bars). Bottom: posterior histograms corresponding to the atmosphere and stellar contamination models. The mean molecular weight ($\mu$) --- a derived property from the atmospheric mixing ratio parameters --- rules out light atmospheres dominated by H$_2$ and He ($\mu \approx 2.3$\,amu). The statistically favored solution is a high mean molecular weight atmosphere dominated by SO$_2$ ($\mu = 64$\,amu).}
\label{fig:poseidon_results}
\end{figure*}

Our POSEIDON retrieval analysis finds a weak statistical preference for the atmosphere model. In terms of the Bayesian evidence, the evidence for the atmosphere model ($\ln \mathcal{Z} = 1895.3$) is higher than both the flat line ($\ln \mathcal{Z} = 1893.0$) and stellar contamination models ($\ln \mathcal{Z} = 1892.5$). Similarly, an atmosphere is preferred by the $\chi^2$ metric (atmosphere: $\chi^2 = 186$ with 207 degrees of freedom; flat line: $\chi^2 = 198$ with 216 degrees of freedom; stellar contamination: $\chi^2 = 197$ with 211 degrees of freedom), but we note that neither the flat line nor the stellar contamination models can be formally rejected given the present data uncertainties. We find that the evidence for an atmosphere arises from multiple small spectral features in \planetname's transmission spectrum that are best fit by SO$_2$ absorption. To quantify the evidence for SO$_2$, we conducted an additional nested Bayesian model comparison by running a further atmosphere model retrieval without SO$_2$ included. We find a Bayes factor of 4.0 (2.2$\sigma$) in favor of an atmosphere including SO$_2$ over alternative atmospheric compositions. We find consistent retrieval results for the coarser \texttt{Eureka!} data bin sizes and for the \texttt{FIREFLy} data reduction, with a similar $\approx$ 2$\sigma$ preference for SO$_2$.

Fig.~\ref{fig:poseidon_results} summarizes our POSEIDON retrieval results. Our \planetname\ transmission spectrum rules out thick low mean molecular weight atmospheres ($\mu > 20.1$\,amu to 2$\sigma$), with a 2$\sigma$ upper limit on the H$_2$ mixing ratio of 24\%. The favored solution is an SO$_2$-rich atmosphere ($\sim$ 100\% SO$_2$), but a wide range of lower SO$_2$ abundances are also consistent with the present observations. Our retrievals additionally disfavor CO$_2$-rich atmospheres (CO$_2$ $<$ 84\% to 2$\sigma$), but the present data do not constrain the N$_2$, CH$_4$, or H$_2$S abundances. We see from Fig.~\ref{fig:poseidon_results} that the evidence for SO$_2$ arises primarily from two absorption bands in the 3.9--4.5~$\mu$m region covered by the NRS2 detector and half an absorption band in the 2.8--3.1~$\mu$m region covered by the NRS1 detector. A wide range of surface pressures are possible ($P_{\rm{surf}} > 10^{-5}$\,bar to 2$\sigma$, i.e. no high-altitude cloud deck is detected), with the maximum posterior densities corresponding to atmospheres with $P_{\rm{surf}} > 1$\,bar. Therefore, we conclude that the most likely explanation for \planetname's transmission spectrum is an SO$_2$-rich atmosphere.

\section{Discussion}
The preference for an SO$_2$ atmosphere over a featureless transmission spectrum in \planetname\ is consistent across independent data reductions and atmospheric retrievals. Moreover, the best-fit atmospheric model matches the predictions from \citet{seligman2024tidal} of a volcanically active planet to surprisingly high fidelity. Six more transits of this planet may add enough signal to provide stronger $\sim5\sigma$ evidence for this atmospheric composition based on these predictions. In this Section, we proceed under the assumption that the \ce{SO2}-dominated scenario is correct.

\subsection{Atmospheric chemistry}
An \ce{SO2}-dominated atmosphere on the warm sub-Earth-sized planet \planetname\ would provide a unique planetary environment to study atmospheric chemistry, lifetime, and implications on the geologic activity. Our self-consistent models indicate that such an atmosphere would have a long photochemical lifetime and no expected haze layer that would interfere with the transmission spectrum. Sulfur hazes, which are self-consistently included in the photochemical model, do not form in this atmosphere because the saturation vapor pressure increases quickly with temperature. The photolysis of \ce{SO2}, which proceeds mainly as photoexcitation instead of direct photodissociation, results in large quantities of \ce{SO3} and gas-phase elemental sulfur (Fig.~\ref{fig_forward}). Elemental sulfur would result in a moderate temperature inversion in the upper atmosphere.  

\subsection{Lifetime of the atmosphere}\label{sec:lifetime}
An important consideration is the lifetime of an \ce{SO2} atmosphere against atmospheric escape. Assuming a present-day XUV flux 0.1 times that of the saturation phase \citep{fromont2024} and a conservative 1\% escape efficiency, we estimate an energy-limited mass-loss rate of $\dot{M}_{\rm escape} \sim 2 \cdot 10^{5}~$kg/s. Without replenishment, a 10-bar atmosphere would be lost in $\sim0.01$ billion years. Therefore it is likely that the atmosphere favored by our observations is in a steady state where the escape is balanced by continuous volcanic outgassing, in that $\dot{M}_{\rm escape} = M_{\rm volc}\cdot x$, where $M_{\rm volc}$ is the extrusive volcanic rate and $x$ is the volatile (i.e., sulfur and carbon) content in the magma. Using the bulk silicate Earth's value for $x$ ($\sim100-200$ ppm, \citealt{gaillard2022}), we estimate an extrusive volcanic rate of $M_{\rm volc}\sim 1-2 \cdot 10^{9}~$kg/s. In comparison, Io's value is $M_{\rm volc, Io}\sim 7 \cdot 10^{6}~$kg/s. This means that, per unit mass, \planetname\ would experience about eight times as much volcanic outgassing and tidal heating as Io. These values can generally be considered as lower limits since we assumed an escape efficiency of only 1\%.

It is worth considering why the planet may have managed to retain some of its volatile inventory despite its proximity to its host star. Assuming an age of 5 billion years \citep{engleguinan2023} and that XUV-driven escape can remove volatiles from the bulk silicate part of the planet without an additional bottleneck, then even a $1\%$ escape efficiency would result in $>2\%$ of the planetary mass lost. For comparison, chondrites have $1-5\%$ bulk sulfur content by mass \citep{alexander2022}. Therefore, it is conceivable that enough sulfur remains. Alternatively, assuming the planet is cold enough to have a solid surface, a lithosphere may have acted as a bottleneck to tamper the escape, trapping the volatiles in the interior, which were eventually volcanically outgassed \citep{kite2020}.

\subsection{Implications on the interior properties of \planetname}
\begin{figure*}
\centering
\includegraphics[width=0.45\linewidth]{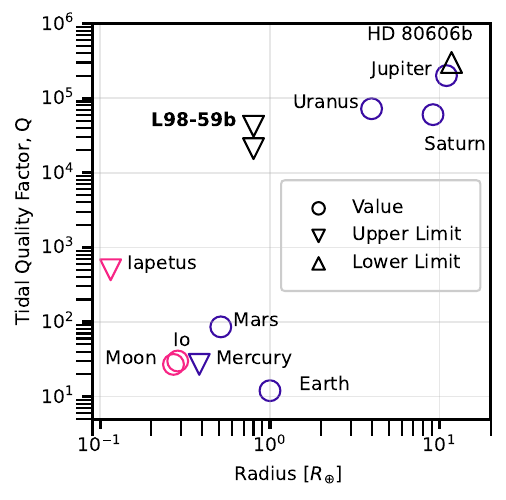}
\includegraphics[width=0.45\linewidth]{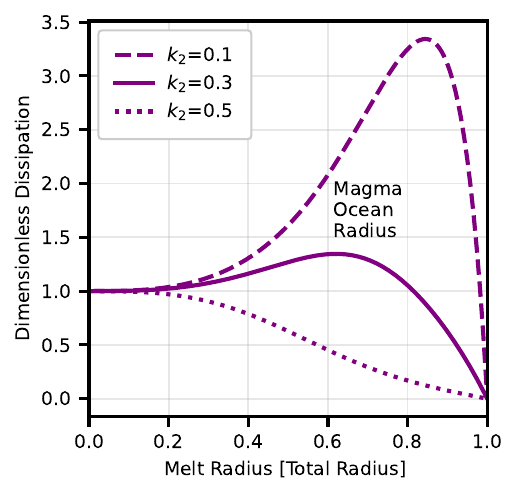}
\caption{The existence of widespread volcanism on \planetname\ --- if caused by the runaway melting mechanism --- provides constraints on the tidal quality factor and size of the subsurface magma ocean. \texttt{Left:} Tidal quality factor versus radius for solar system planets and satellites along with the constraint for \planetname\ given by Eq.~\ref{Qscale} (i.e., $Q<1400~Q_{\rm Io}$, see Appendix~\ref{app:love}) and by assuming that volcanic outgassing must balance atmospheric escape ($Q<700~Q_{\rm Io}$, see Section~\ref{sec:lifetime}). \textit{Right:} Volume integrated tidal heating for a composite body consisting of a liquid interior surrounded by a solid mantle as a function of subsurface magma ocean radius, normalized to the case of no magma ocean. The location where the derivative of this function with respect to the melt radius is zero indicates the approximate equilibrium value for the runaway melting process. }\label{fig:tidalq_meltfraction}
\end{figure*}

The presence of volcanic activity on \planetname\ can be used to infer geophysical and geochemical properties about the planet's interior. The existence of widespread volcanic activity ---  if a product of the runaway melting mechanism outlined by \citep{Peale1978} and applied to Io \citep{Peale1979}-- provides a constraint on the tidal quality factor, $Q$, of the planet (see Appendix~\ref{app:love}). By assuming that \planetname\ experiences as much or more tidal heating per unit mass as Io yields the constraint that $Q_{\rm{\planetname}}\lesssim1400\,Q_{Io}$. In Fig.~\ref{fig:tidalq_meltfraction}
we show where this constraint places \planetname\ in comparison to planets and satellites for which Q has been measured \citep{Murray1999,Laughlin2009,deWit2016,Lainey2009}. The runaway melting should be operating if \planetname\ has a quality factor in the range of all of those measured in rocky planets and satellites in the solar system (Fig.~\ref{fig:tidalq_meltfraction}). Moreover, if the runaway melting mechanism is operating, then assuming a love number $k_2\sim0.1-0.5$ provides an approximate size of the subsurface magma ocean. In the right panel of Fig.~\ref{fig:tidalq_meltfraction}, we show that the runaway melting mechanism should produce a subsurface magma ocean of $R_{m}\sim0.6-0.9R_{p}$.

A volcanic atmosphere rich in sulfur dioxide on \planetname\ would be indicative of an oxidized mantle with an oxygen fugacity larger than +2.7 log units relative to the iron-wüstite (IW) buffer \citep{liggins2022}. Besides \ce{SO2}, these outgassed atmospheres typically contain high abundances of \ce{H2O}, \ce{CO2}, and \ce{H2S}, although it is likely that \planetname\ has experienced rapid water loss \citep{fromont2024}. In our Bayesian retrievals, we tested models that included these additional gases, but we could not derive meaningful constraints.

\subsection{The cosmic shoreline}
When the planets and moons in the Solar System are arranged on an insolation versus escape velocity diagram, there is a “cosmic shoreline” that separates worlds that have an atmosphere from those that do not \citep{zahnle2017}. \planetname\ falls on the side of the shoreline where we would expect it to be predominantly airless \citep{pidhorodetska2021}. If an \ce{SO2} atmosphere is indeed present on \planetname, it would support the idea that volcanic activity can replenish the atmosphere that was perhaps once lost on this and other similar planets.

Hints of a sulfur atmosphere on the more massive L\,98-59\,d \citep{Gressier2024,Banerjee2024}, a target that may also be undergoing mantle melting \citep{seligman2024tidal}, could signal widespread volcanism in the \starname\ system. However, the mass and radius of planet d are inconsistent with a purely rocky composition \citep{demangeon2021l9859b,luquepalle2022}. JWST observations of tidally heated planets around M dwarfs may help us elucidate how widespread this mechanism is and place constraints on the bulk geophysical properties of these worlds.

\section{Conclusions}
We observed four transits of the sub-Earth \planetname\ using JWST/NIRSpec G395H to search for a volcanically outgassed atmosphere. Overall, our analysis finds that while frequentist metrics like the p-value cannot reject the null hypothesis of a flat line or bare rock model ($p=0.81$), an atmospheric model with SO$_2$ absorption can also explain the observations given their signal-to-noise. A model comparison of diverse atmospheric models against the bare rock model results in tantalizing preferences for the atmospheric scenario at the $\gtrsim3\sigma$ level. These model comparisons have their own weaknesses and can be misleading under pathological scenarios \citep{Welbanks2022, Welbanks2023}. The use of other model comparison metrics such as LOO-CV also suggest that the observations are compatible with an SO$_2$-rich model and highlight that most of the preference for this atmospheric model comes from regions where SO$_2$ absorption is the dominant source of opacity.

Assuming that \planetname\ hosts an \ce{SO2}-dominated atmosphere driven by widespread volcanism, we can infer geophysical and geochemical properties about the interior of the planet. For example, such an atmosphere would suggest the mantle is oxidized, with an oxygen fugacity of $f\rm{O}_2>IW+2.7$. By equaling an estimate of the energy-limited mass loss rate to the volcanic outgassing rate, we calculate that \planetname\ must experience at least eight times as much volcanism and tidal heating as Io. If volcanism is driven by runaway melting of the mantle \citep{Peale1978,Peale1979,seligman2024tidal}, our detailed interior modeling indicates that \planetname\ must host a subsurface magma ocean extending up to $\sim 0.6-0.9~R_p$.

\section*{Data availability}
The JWST NIRSpec data used in this work will be publicly available at the end of the one-year exclusive access period in the Mikulski Archive for Space Telescopes (MAST): \dataset[10.17909/3g7x-b466]{https://dx.doi.org/10.17909/3g7x-b466}. The transmission spectra presented in this work are available on Zenodo: \dataset[10.5281/zenodo.14676142]{https://dx.doi.org/10.5281/zenodo.14676142}

%% IMPORTANT! The old "\acknowledgment" command has be depreciated. It was
%% not robust enough to handle our new dual anonymous review requirements and
%% thus been replaced with the acknowledgment environment. If you try to 
%% compile with \acknowledgment you will get an error print to the screen
%% and in the compiled pdf.
%% 
%% Also note that the akcnowlodgment environment does not support long amounts of text. If you have a lot of people and institutions to acknowledge, do not use this command. Instead, create a new \section{Acknowledgments}.

\vspace{5mm}
We thank the anonymous referee for their thoughtful and insightful report on this work. This work is based in part on observations made with the NASA/ESA/CSA James Webb Space Telescope. The data were obtained from the Mikulski Archive for Space Telescopes at the Space Telescope Science Institute, which is operated by the Association of Universities for Research in Astronomy, Inc., under NASA contract NAS5-03127. These observations are associated with program \#3942. Support for program \#3942 was provided by NASA through a grant from the Space Telescope Science Institute under NASA contract NAS5-03127. Part of this research was carried out at the Jet Propulsion Laboratory, California Institute of Technology, under a contract with the National Aeronautics and Space Administration (80NM0018D0004). Part of the high-performance computing resources used in this investigation were provided by funding from the JPL Information and Technology Solutions Directorate. R.J.M. is supported by NASA through the NASA Hubble Fellowship grant HST-HF2-51513.001, awarded by the Space Telescope Science Institute, which is operated by the Association of Universities for Research in Astronomy, Inc., for NASA, under contract NAS 5-26555. D.Z.S. is supported by an NSF Astronomy and Astrophysics Postdoctoral Fellowship under award AST-2303553. This research award is partially funded by a generous gift of Charles Simonyi to the NSF Division of Astronomical Sciences. The award is made in recognition of significant contributions to Rubin Observatory’s Legacy Survey of Space and Time.

%% To help institutions obtain information on the effectiveness of their 
%% telescopes the AAS Journals has created a group of keywords for telescope 
%% facilities.
%
%% Following the acknowledgments section, use the following syntax and the
%% \facility{} or \facilities{} macros to list the keywords of facilities used 
%% in the research for the paper.  Each keyword is check against the master 
%% list during copy editing.  Individual instruments can be provided in 
%% parentheses, after the keyword, but they are not verified.

%% Similar to \facility{}, there is the optional \software command to allow 
%% authors a place to specify which programs were used during the creation of 
%% the manuscript. Authors should list each code and include either a
%% citation or url to the code inside ()s when available.

%% Appendix material should be preceded with a single \appendix command.
%% There should be a \section command for each appendix. Mark appendix
%% subsections with the same markup you use in the main body of the paper.

%% Each Appendix (indicated with \section) will be lettered A, B, C, etc.
%% The equation counter will reset when it encounters the \appendix
%% command and will number appendix equations (A1), (A2), etc. The
%% Figure and Table counter will not reset.

\appendix
\setcounter{figure}{0}
\renewcommand{\thefigure}{A\arabic{figure}}
\setcounter{table}{0}
\renewcommand{\thetable}{A\arabic{table}}

\section{Data Reduction Supplemental information}
\subsection{Comparison of the transmission spectra across different reductions and lightcurve fitting strategies}\label{app:comparison}
\begin{figure*}[ht]
\centering
\begin{minipage}{0.49\textwidth}
    \centering
    \includegraphics[width=\linewidth]{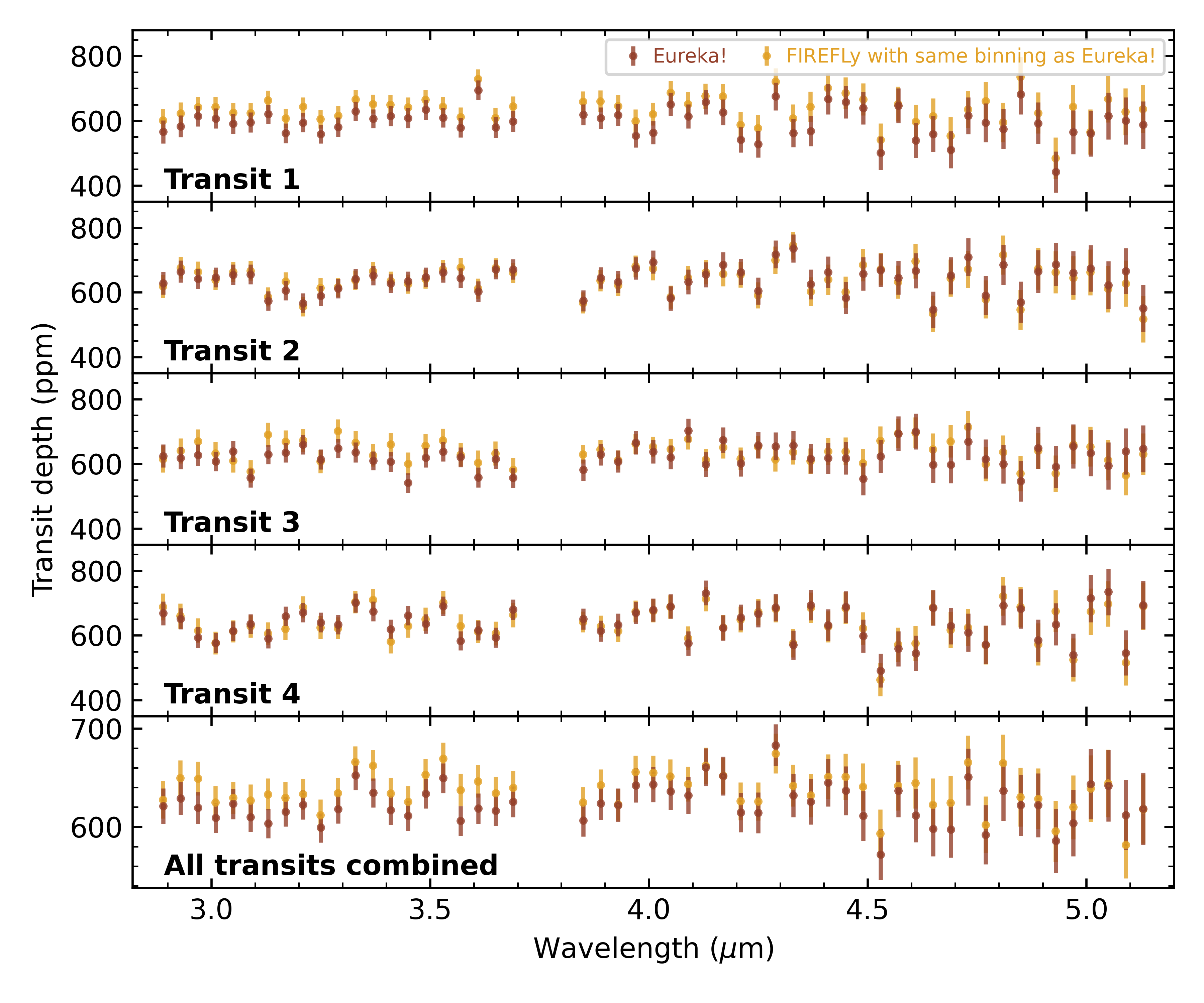}
    \label{transmission_spectra_comparison_reductions}
\end{minipage}\hfill
\begin{minipage}{0.49\textwidth}
    \centering
    \includegraphics[width=\linewidth]{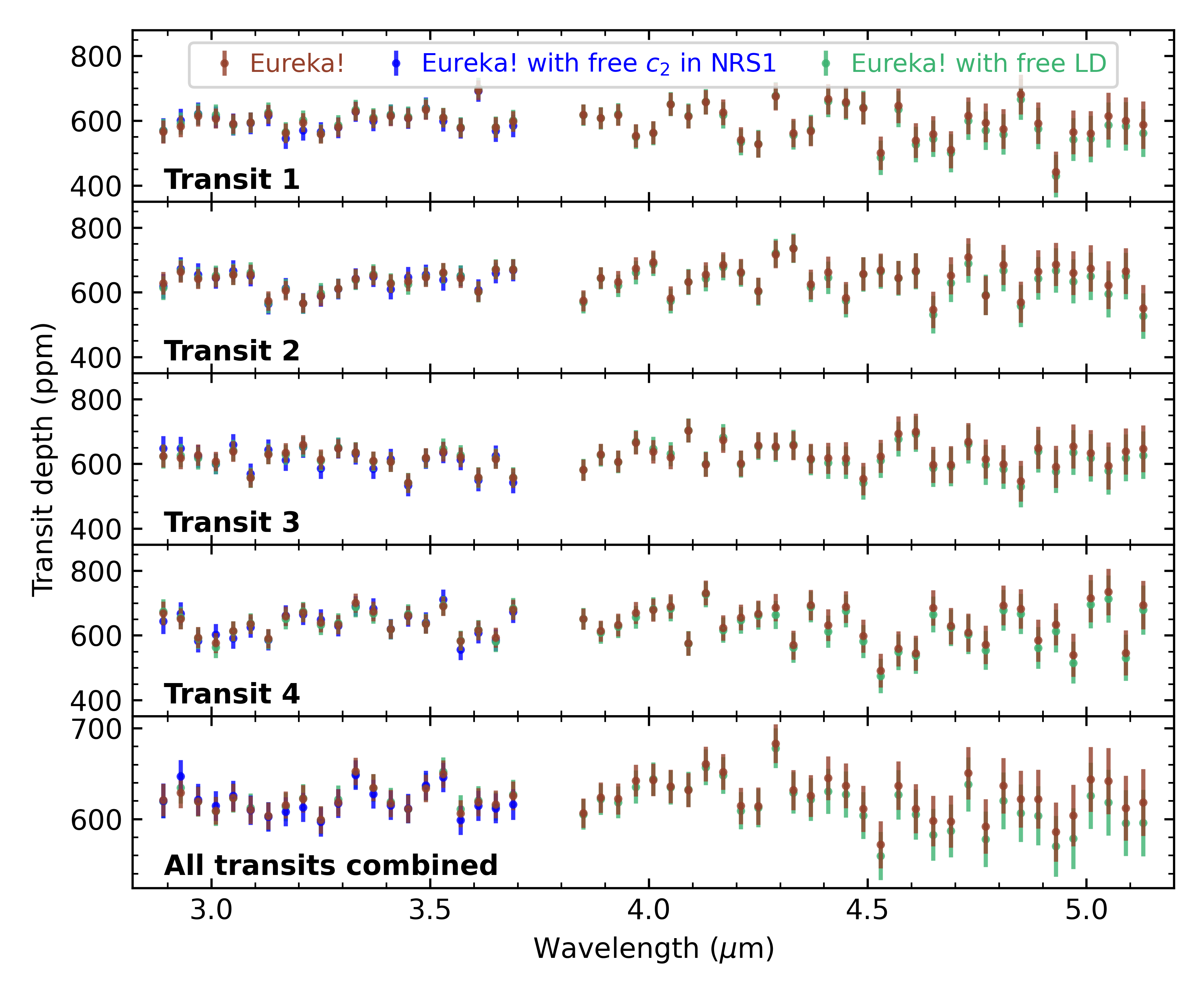}
    \label{transmission_spectra_comparison_eurekaonly}
\end{minipage}
\caption{The transmission spectra of \planetname\ using independent pipelines (left) and, in the case of \texttt{Eureka!}, different lightcurve fitting strategies (right). Brown is our standard \texttt{Eureka!} reduction obtained when we binned the lightcurves to $\Delta \lambda = 0.04 ~ \mu$m, and yellow is the \texttt{FIREFLy} reduction with the same binning (the retrievals on the FIREFLy data used the equal-counts-per-bin reduction, not shown here). Blue is the same reduction as brown, but allowing for the quadratic term of the polynomial ($c_2$, only used in NRS1) to be freely fit in each channel instead of fixing it to the value derived from the white lightcurve. Green is the same reduction as brown, but allowing for limb darkening (LD) to be freely fit in each channel.}
\label{fig:combined_transmission_spectra}
\end{figure*}

As shown in Fig.~\ref{transmission_spectra_comparison_reductions},
the independent reductions produced consistent transmission spectra despite their different strategies to model the systematics and fit the lightcurves. Similarly, freeing the quadratic term of the polynomial or the limb-darkening coefficients in the \texttt{Eureka!} lightcurve fits did not significantly affect the transmission spectra.

\subsection{Allan deviation plots}\label{app:allan}
\begin{figure*}
\centering
\includegraphics[width=0.9\textwidth]{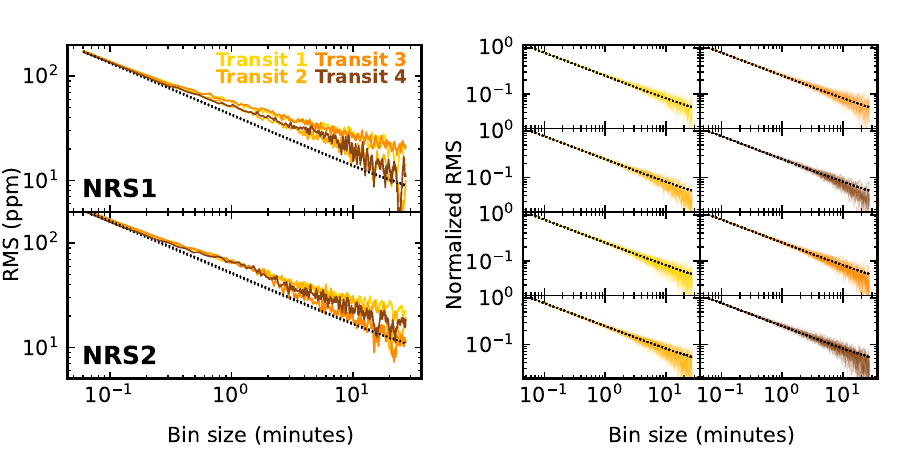}
\caption{Allan deviation plots. \textit{Left:} Allan deviation plots for the \texttt{Eureka!} white lightcurves of both detectors. \textit{Right:} Allan deviation plots for the \texttt{Eureka!} spectroscopic lightcurves. We present the reduction with the $\Delta\lambda = 0.04~\mu$m binning.}\label{fig:allan}
\end{figure*}

In Fig.~\ref{fig:allan}, we present the Allan deviation plots for the white and spectroscopic lightcurves.
The Allan deviation plots for the white lightcurves show some correlated noise with timescales of $\sim 0.3-5$ minutes, typical of the thermal cycling of heaters in the ISIM Electronics Compartment \citep{rigby2023}. Some of them also show time-correlated noise with lower frequencies, possibly driven by the low number of groups per integration. The spectroscopic lightcurves do not show evidence for correlated noise.

\subsection{Limb-darkening coefficients}
\begin{figure}
\centering
\includegraphics[width=\linewidth]{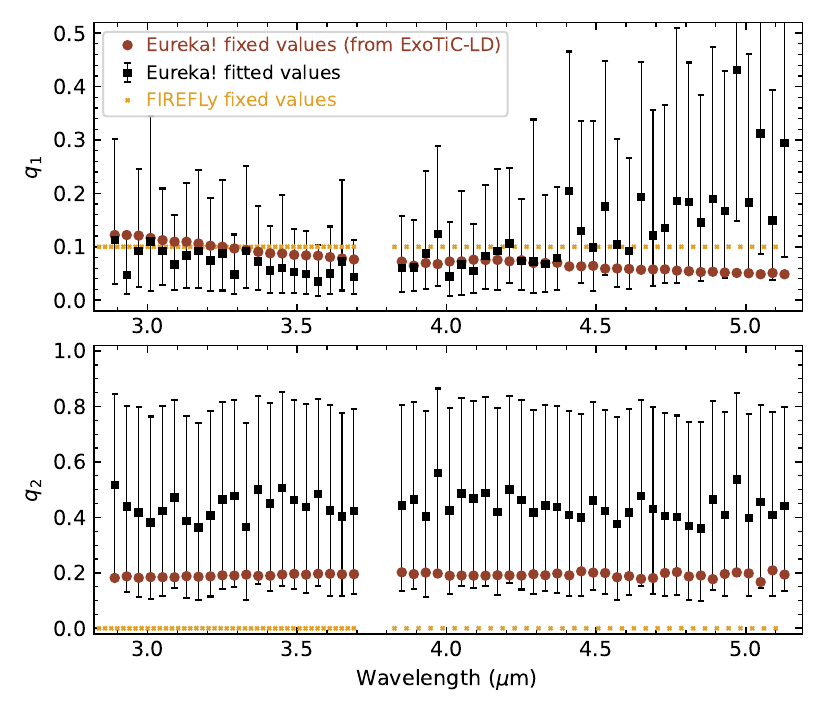}
\caption{Different choices of limb-darkening coefficients. The \texttt{Eureka!} reduction fixed the 
quadratic limb darkening coefficients $u_1$ and $u_2$ to those calculated with \texttt{ExoTiC-LD} (in the plot we show $q_1=(u_1+u_2)^2$ and $q_2=0.5 u_1(u_1+u_2)^{-2}$ instead, \citealt{kipping2013limbdark}), while \texttt{FIREFLy} opted for fixing $q_1=0.1$ and $q_2=0$. We also present the values from an alternative \texttt{Eureka!} reduction in which $q_1$ and $q_2$ were freely fitted in each spectroscopic lightcurve. The uncertainties correspond to the 16$^{\textup{th}}$ and 84$^{\textup{th}}$ percentiles in all samples in each channel.}\label{fig:ld_comparison}
\end{figure}

Fig.~\ref{fig:ld_comparison}
presents the limb darkening coefficients used in the different reductions. Our standard \texttt{Eureka!} reduction fixed the limb-darkening coefficients to those calculated with \texttt{ExoTiC-LD} \citep{grantwakeford2022exoticld}, and FIREFLy fixed them to $q_1=0.1$ and $q_2=0$. When we fit for the limb-darkening coefficients instead of fixing them, we find that the $q_1$ values cluster around 0.1, and $q_2$ is essentially unconstrained, with posteriors that are flat from 0 to 1 at all NIRSpec G395H wavelengths. $q_2$ has flat posteriors at all wavelengths.

\subsection{Best-fit orbital parameters from the white lightcurves}\label{app:bestfit_values}
Table~\ref{tab2} shows the best-fit orbital parameters from the white lightcurve fits. There are significant differences in the values of $R_p/R_s$ across different transits. Within a single transit, the fitted $R_p/R_s$ values also differ significantly by reduction, pointing to correlated noise, not stellar inhomogeneities, as the driver of these discrepancies. We tried a similar approach to \texttt{FIREFLy!} of iteratively fixing $i$ and $a/Rs$, but the significant offsets remain. This may be due to the  correlated noise in the white lightcurves. Since the white lightcurve fits are only used to fix the orbital parameters and the quadratic polynomial terms of the systematics during the spectroscopic fits, increasing the error bars of the white lightcurves to account for red noise would not have a significant impact on the final transmission spectra. However, as part of our retrievals with Aurora, we ran a test to study the influence of offsets in $R_p/R_s$ across different transits. 

\begin{table*}
\centering
\small
\begin{tabular}{c|cc|c|c|c|c}
\hline
\hline
              Reduction    & \multicolumn{2}{c|}{Data Set}                     &        $T_0$ (BMJD$_\textup{TDB}$)           &         $i$ ($^\circ$)          &   $a/R_s$                & $R_p/R_s$  \\ \hline
\multirow{8}{*}{\texttt{Eureka!}} & \multicolumn{1}{l|}{\multirow{2}{*}{Transit 1}}  & NRS1 & $60339.398572^{+5.7e-5}_{-5.5e-5}$   & $88.76^{+0.58}_{-0.62}$  & $16.73^{+0.85}_{-1.20}$   & $0.02455^{+0.00023}_{-0.00021}$  \\  
                  & \multicolumn{1}{l|}{}                  & NRS2 & $60339.398562^{+5.8e-5}_{-5.6e-5}$   & $89.27^{+0.50}_{-0.65}$  & $17.34^{+0.40}_{-0.98}$    &  $0.02437^{+0.00021}_{-0.00022}$   \\  
                  & \multicolumn{1}{l|}{\multirow{2}{*}{Transit 2}} & NRS1 & $60343.905097^{+6.2e-5}_{-6.2e-5}$   & $87.20^{+0.47}_{-0.42}$  & $13.51^{+1.01}_{-0.86}$     & $0.02568^{+0.00024}_{-0.00025}$   \\  
                  & \multicolumn{1}{l|}{}                  & NRS2 & $60343.904827^{+6.2e-5}_{-6.2e-5}$   & $87.97^{+0.51}_{-0.49}$  & $15.16^{+1.06}_{-1.05}$     &  $0.02569^{+0.00024}_{-0.00022}$     \\  
                  & \multicolumn{1}{l|}{\multirow{2}{*}{Transit 3}} & NRS1 & $60346.157872^{+5.8e-5}_{-6.1e-5}$   & $88.08^{+0.55}_{-0.52}$  & $15.43^{+1.11}_{-1.12}$     & $0.02497^{+0.00024}_{-0.00022}$ \\  
                  & \multicolumn{1}{l|}{}                  & NRS2 & $60346.157945^{+6.5e-5}_{-6.4e-5}$    & $89.27^{+0.49}_{-0.58}$  & $17.47^{+0.41}_{-0.91}$     &  $0.02511^{+0.00020}_{-0.00021}$     \\   
                  & \multicolumn{1}{l|}{\multirow{2}{*}{Transit 4}} & NRS1 & $60359.676715^{+5.2e-5}_{-4.7e-5}$   & $88.83^{+0.54}_{-0.53}$  & $16.91^{+0.75}_{-1.01}$     &  $0.02520^{+0.00020}_{-0.00019}$     \\  
                  & \multicolumn{1}{l|}{}                 & NRS2 & $60359.676755^{+5.2e-5}_{-5.5e-5}$    & $88.60^{+0.59}_{-0.58}$  & $16.34^{+0.93}_{-1.15}$     &  $0.02550^{+0.00022}_{-0.00020}$     \\ \hline
\multirow{8}{*}{\texttt{FIREFLy}} & \multicolumn{1}{l|}{\multirow{2}{*}{Transit 1}} & NRS1 & \multirow{2}{*}{$60339.398605\pm2.0e-5$} & \multirow{8}{*}{$88.76\pm0.02$} & \multirow{8}{*}{$16.72\pm0.30$} & $0.02591\pm0.00029$ \\   
                  & \multicolumn{1}{l|}{}                  & NRS2 &                   &                   &                   & $0.02543\pm0.00045$ \\   
                  & \multicolumn{1}{l|}{\multirow{2}{*}{Transit 2}} & NRS1 & \multirow{2}{*}{$60343.904832\pm2.0e-5$} &                   &                   & $0.02445\pm0.00029$ \\   
                  & \multicolumn{1}{l|}{}                  & NRS2 &                   &                   &                   & $0.02449\pm0.00047$ \\   
                  & \multicolumn{1}{l|}{\multirow{2}{*}{Transit 3}} & NRS1 & \multirow{2}{*}{$60346.157946\pm2.0e-5$} &                   &                   & $0.02574\pm0.00025$ \\   
                  & \multicolumn{1}{l|}{}                  & NRS2 &                   &                   &                   & $0.02488\pm0.00018$ \\   
                  & \multicolumn{1}{l|}{\multirow{2}{*}{Transit 4}} & NRS1 & \multirow{2}{*}{$60359.676627\pm2.0e-5$} &                   &                   & $0.02591\pm0.00023$ \\   
                  & \multicolumn{1}{l|}{}                  & NRS2 &                   &                   &                   & $0.02362\pm0.00039$ \\ \hline
\end{tabular}
\vspace{5pt}
\caption{Best-fit Orbital Parameters and 1$\sigma$ Uncertainties from Fitting the White Lightcurves. \texttt{FIREFLy} fits the final white lightcurves using a weighted mean of $T_0$, $a/R_s$, and $b$ (converted into $i$ here), from all eight lightcurves. }\label{tab2}
\end{table*}

\section{Modeling an \ce{so2}-dominated atmosphere with EPACRIS}\label{app:epacris}
We first calculated the pressure-temperature (\textit{P--T}) profile of an \ce{SO2}-only atmosphere under radiative-convective equilibrium using the climate module of the ExoPlanet Atmospheric Chemistry \& Radiative Interaction Simulator (EPACRIS-Climate, Scheucher et al., in preparation). EPACRIS implemented the two-stream method of \citet{heng2018radiative} to calculate the radiative fluxes, and solve the temperature profile that minimizes the net radiative fluxes in all atmospheric layers using a Newton method. EPACRIS incorporated both dry and moist adiabatic adjustments following \citet{graham2021multispecies}, and in the cases studies here, only dry adiabatic adjustments were triggered. We assume a Bond albedo of zero and consider an internal heat flux that corresponds to $1\times$ and $10\times$ the insolation on \planetname, following the tidal heating rate estimates by \citet{seligman2024tidal} and covering the range of $Q\sim3-30$ \citep{tobie2019tidalheating}. For simplicity, we assume the atmosphere to be 1 bar, and the surface to be fully absorptive at all wavelengths.

After obtaining the initial \textit{P--T} profile, we ran our photochemical model (described below) to trace the formation and evolution of molecular species other than \ce{SO2}. Then, we used the steady-state mixing ratio profiles produced by the photochemical model to recalculate the \textit{P-T} profile, now incorporating the opacities of not only \ce{SO2} but also the photochemically produced species. This process was iterated three times, when the \textit{P-T} profile and the corresponding steady-state atmospheric mixing ratio profiles no long change. In these calculations, the infrared opacities of \ce{SO2} and SO$_3$ are obtained from the DACE database \citep{grimm2021helios}. The UV and visible-wavelength opacities of sulfur allotropes, oxygen, and sulfur oxides are obtained from the compilation in \citet{hu2012photochemistry}, and \citet{burkholder1997uv} and \citet{hintze2003vibrational} for \ce{SO3}. For the lack of available data, we adopt the UV and visible cross sections of \ce{S4} for \ce{S8} \citep{billmers1991ultraviolet}.

We performed 1D photochemical kinetic-transport atmospheric modeling using the chemistry module of EPACRIS, with the chemical network in \citet{hu2012photochemistry} and updated by \citet{Hu_2021} and \citet{wogan2024jwst}. Similarly to \citet{Hu_2021}, the eddy diffusion coefficient profile was assumed to be 10$^3$~cm$^2$~s$^{-1}$ in the convective part of the bottom atmosphere, and vary as $n^{-1/2}$ in the radiative part of the atmosphere, $n$ being the number density. We adopted the panchromatic stellar spectrum of GJ~176 from the MUSCLES survey \citep{Loyd_2016} as a proxy for \starname. For simplicity, we assumed a constant mixing ratio of \ce{SO2} at the lower boundary and zero flux for all other species at the boundaries. We did not observe the formation of sulfur aerosols in these models.

After the models had reached the steady state, we computed the synthetic transmission spectra of \planetname\ based on the molecular mixing ratio profiles of corresponding scenarios, using the transmission spectra generation module of EPACRIS \citep{Hu_2013}. 

The formation of \ce{SO3} is driven by the photoexcitation of \ce{SO2} in the middle atmosphere: \ce{SO2 + hv -> ^1SO2} and \ce{SO2 + hv -> ^3SO2}. The excited states \ce{SO2} have short lifetimes in the atmosphere, and \ce{^1SO2} can also become \ce{^3SO2}. A small fraction of \ce{^1SO2} and \ce{^3SO2} participate in subsequent chemical reactions. \ce{SO3} is formed by \ce{^1SO2 + SO2 -> SO + SO3} and \ce{^3SO2 + SO2 -> SO + SO3} \citep{turco1982stratospheric}. The loss of \ce{SO3} mainly proceeds by \ce{SO + SO3 -> SO2 + SO2}, which maintains the steady-state mixing ratio of \ce{SO3}. Elemental sulfur is produced by \ce{SO + SO -> S + SO2}, and some of the S produced self-combine to form \ce{S2}. The net loss of elemental sulfur mainly proceeds via \ce{S + O2 -> SO + O} and \ce{S2 + O -> SO + S}. Some of the \ce{S2} further polymerize to form \ce{S4} and eventually \ce{S8}, which is assumed as the terminal species of sulfur allotropes and can thus accumulate in the atmosphere. The details of the chemical reactions involve sulfur allotropes are uncertain, which may impact the final abundance of \ce{S8}. The photoexcitation, rather than the direct photodissociation of \ce{SO2}, dominates in this atmosphere because the direct photodissociation of \ce{SO2} requires photons with $\lambda<220\ \mu$m, while the photoexcitation can be driven by the photons with wavelengths up to 400 nm and these photons can penetrate deeper into the middle atmosphere \citep{bogumil2003measurements,whitehill2015so}.

\section{Retrieval scenarios from \texttt{ExoTR}}\label{app:exotr}
In Table~\ref{tab:exotr_setup}, we present the priors assigned to each free parameter in \texttt{ExoTR}, 
\begin{table}
    \begin{tabular}{lr}\hline\hline
    \textbf{Parameter}  & \textbf{Prior} \\
    \hline\hline
        Datasets offsets [ppm] & $\mathcal{U}$(-100, 100) \\
        Planetary radius [$R_{\oplus}$] & $\mathcal{U}$(0.5, 2)$\times$ $R_p(^2)$ \\
        Planetary temperature [K] & $\mathcal{U}$(100, 1000) \\
        Cloud top [Pa]  & $\mathcal{LU}$(0.0, 9.0)\\
        VMR H$_2$O  & CLR$(-12, 0)(^1)$ \\
        VMR CH$_4$  & CLR$(-12, 0)(^1)$  \\
        VMR H$_2$S  & CLR$(-12, 0) (^1)$ \\
        VMR \ce{SO2} & CLR$(-12, 0)(^1)$  \\
        VMR \ce{SO3} & CLR$(-12, 0)(^1)$  \\
        VMR CO$_2$  & CLR$(-12, 0)(^1)$  \\
        VMR N$_2$ (derived)  & CLR$(-12, 0)(^1)$ \\
        Heterogeneity fraction & $\mathcal{U}$(0.0 - 0.5) \\
        Heterogeneity temperature [K] & $\mathcal{U}$(0.5, 1.2) $\times$ $T_{\rm eff}$\\
        Stellar temperature [K]  & $\mathcal{N}$(3415, 135)$(^2)$ \\
    \hline
    \end{tabular}
    \vspace{5pt}
    \caption{\texttt{ExoTR} parameters and prior probability distributions used in the atmospheric retrievals. $\mathcal{U}(a,b)$ is the uniform distribution between values $a$ and $b$, $\mathcal{LU}(a,b)$ is the log-uniform (Jeffreys) distribution between values $a$ and $b$, and $\mathcal{N}(\mu,\sigma^2)$ is the normal distribution with mean $\mu$ and variance $\sigma^2$. NOTE - $(^1)$ \citet{damiano2021prior}, $(^2)$ \citet{demangeon2021l9859b}. \label{tab:exotr_setup}}
\end{table}
and in Table~\ref{tab:exotr_res}, we show the results for the retrieval scenarios that were explored.
\begin{table*}
    \begin{tabular*}{\textwidth}{@{\extracolsep\fill}cccccc}\hline\hline
    \textbf{Scenario \#} & \textbf{Description} & \textbf{ln(EV)} & \textbf{$\sigma$ baseline} & $\mathbf{\chi^2/\nu}$ & $\mathbf{\chi^2_\nu}$\\
    \hline\hline
            JWST & & & & & \\
        \textbf{1.} & 100\% \ce{SO2} & 1897.44 & $3.53$ & $189.48/216$ & 0.88\\
        \textbf{2.} & $T_p$, H$_2$O, CH$_4$, H$_2$S, \ce{SO2} (fill), SO$_3$, CO$_2$, NH$_3$, and CO & 1896.77 & $3.33$& $188.87/208$ & 0.91\\
        \textbf{3.} & \ce{SO2} (fill), and SO$_3$ & 1896.24 & $3.17$ & $189.41/215$ & 0.88\\
        \textbf{4.} & \ce{SO2} (fill), SO$_3$, and clouds & 1896.16 & $3.15$ & $189.99/214$ & 0.89\\
        \textbf{5.} & $T_p$, N$_2$, H$_2$O, CH$_4$, H$_2$S, and CO$_2$ & 1892.66 & $<1$ & 197.47/211 & 0.94\\
        \textbf{6.} & Bare Rock & 1892.66 & $-$ & 197.84/216 & 0.92\\
        \textbf{7.} & H$_2$S, and CO$_2$ & 1881.76 & $-$ & 221.51/215 & 1.03\\
    \hline
             HST + JWST & & & & & \\
         \textbf{8.}& $T_p$, N$_2$, H$_2$O, CH$_4$, H$_2$S, \ce{SO2} (fill), SO$_3$, CO$_2$, and clouds & 2063.91 & $3.37$ & $206.39/225$ & 0.92\\
        \textbf{9.} & $T_p$, N$_2$, H$_2$O, CH$_4$, H$_2$S, \ce{SO2}, CO$_2$, and stellar heterogeneity & 2061.40 & $2.43$ & $210.06/224$ & 0.94\\
        \textbf{10.} & $T_p$, N$_2$, H$_2$O, CH$_4$, H$_2$S, CO$_2$, and clouds & 2060.75 & $2.14$ & $212.44/227$ & 0.94\\
         \textbf{11.} & Bare Rock & 2059.67 & $-$ & $214.61/233$ & 0.92\\

    \hline
    \end{tabular*}
    \vspace{5pt}
    \caption{Retrieval scenarios explored with \texttt{ExoTR}. The $\sigma$ baseline is the sigma significance of the scenario when compared with the baseline scenario, i.e., the bare rock scenario. We also present the $\chi^2$, number of degrees of freedom ($\nu$) and reduced $\chi^2$ values ($\chi^2_\nu$) that corresponds to the maximum a posteriori solution of each scenario. All scenarios fit offsets between datasets and the planet radius R$_p$.} \label{tab:exotr_res}
\end{table*}

\section{\ce{SO2}/\ce{CO2} abundance ratio}
\begin{figure}
\centering
\includegraphics[width=\linewidth]{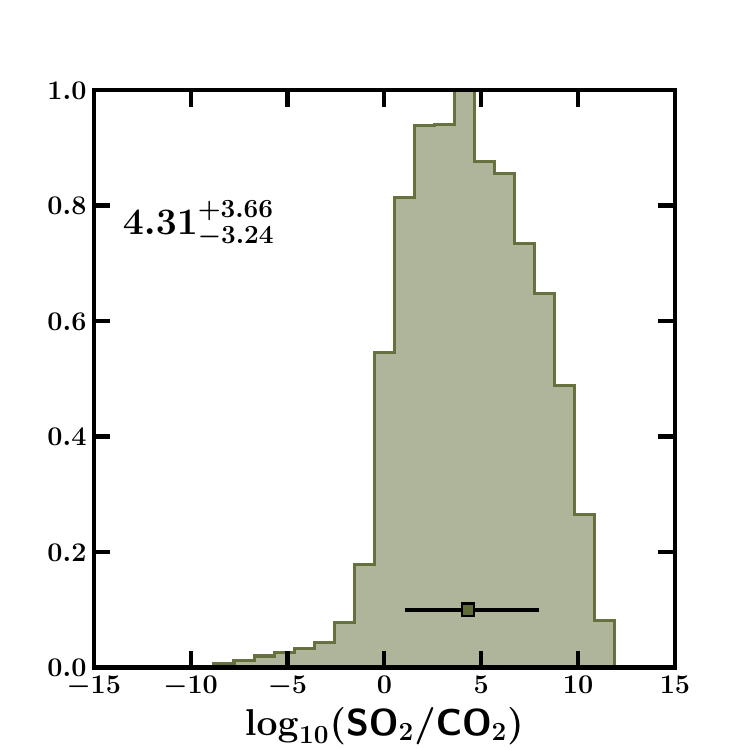}
\caption{\ce{SO2}/\ce{CO2} abundance ratio retrieved with Aurora.}\label{fig:ratio_so2_co2}
\end{figure}

Fig.~\ref{fig:ratio_so2_co2} shows the \ce{SO2}/\ce{CO2} abundance ratio from the 14-parameter retrieval with Aurora.

\section{Calculation of the Love number and magma ocean radius of \planetname}\label{app:love}
The confirmation of an \ce{SO2} enriched atmosphere would be compelling evidence --- although not definitive --- for active and widespread surface volcanism on \planetname. The presence of volcanic activity would, in turn, be indicative that similar tidal heating mechanisms operate in the \starname\ stellar and planetary system as in the Jupiter-Io system. This could potentially lead to an approximate  constraint on the tidal quality factor of \planetname. As a zeroth-order approximation, we assume that the rate of tidal heating per unit mass is approximately equal in both Io and \planetname. The rate of tidal heating, $\dot{E}_\textrm{Heat}$ is given by
\begin{equation}\label{eq:scaled_heating_earth}
\begin{split}
     \dot{E}_\textrm{Heat} =\,\big(\, 3.4\times10^{25}\, \textrm{erg\,s}^{-1}\,\big)\,\,\bigg( \frac{P}{1\textrm{d}} \bigg)^{-5}\\ \,\bigg( \frac{R_{\rm p}}{R_\oplus}\bigg)^5 \bigg(\frac{e}{10^{-2}}\bigg)^2\,\bigg(\frac{Q}{10^{2}}\bigg)^{-1}  \,. 
     \end{split}
\end{equation}
In Equation (\ref{eq:scaled_heating_earth}) $P$ is the orbital period, $R_{\rm P}$ is the planet/satellite radius, $e$ is the orbital eccentricity and $Q$ is the tidal quality factor. By plugging in the relevant values for Io and for \planetname\ and assuming that both objects have the same bulk density, we calculate the relationship between the rate of tidal heating between both bodies and their quality factors,
\begin{equation}\label{eq:edot_io_l98}\dot{E}_{\rm{\planetname}}\gtrsim3\times10^4\,\bigg(\frac{Q_{\rm{\planetname}}}{Q_{\rm{Io}}}\,\bigg)\,\dot{E}_{\rm{Io}}\,.
\end{equation}
To calculate Equation (\ref{eq:edot_io_l98}) we assume that \planetname\ receives comparable or more tidal heating per unit mass as Io. This yields a rough constraint on the quality factor, assuming that both bodies have the same bulk density,
\begin{equation}\label{Qscale}
    Q_{\rm{\planetname}} \lesssim 1400 \,Q_{\rm{Io}}\,.
\end{equation}

The existence of volcanic activity could also provide insights into the  interior structure of \planetname. The following argument relies on the  assumption that the runaway melting mechanism is operating. However, it should be noted that more sophisticated models exist for the volcanism on Io \citep{Lopes2007,Keane2023}. Here we review the calculations presented in \citet{seligman2024tidal} and apply them to the case of \planetname.

The Love number, $k_2$, for a composite planet consisting of a melted interior and a rocky mantle (with rigidity $\mu$), is given by
\begin{equation}\label{eq:k2_2layer}
    k_{2} = \frac{3}{2}\bigg(1+\mathcal{Z}(\xi) \frac{\mu}{\rho g R_{\rm p}}\bigg)^{-1}\,.
\end{equation}
The parameter $\xi$ is the radius of the  melted interior  compared to the total radius, and $\rho$ is the bulk density of the planet. The function $\mathcal{Z}(\xi) $ is given by
\begin{equation}\label{eq:zliq}\begin{split}
    \mathcal{Z}(\xi) = 12 \\\,\bigg(\,\frac{19-75\xi^3+112\xi^5-75\xi^7+19\xi^{10}}{24+40\xi^3-45\xi^7-19\xi^{10}}\,\bigg)\,.
    \end{split}
\end{equation}
which was derived by \citet{Beuthe2013}. 
We assume that $k_2=0.1,0.3$ or $0.5$ for the planet \citep{tobie2019tidalheating}. Using these values for $k_2$, we solve for the amount of energy dissipated for the melted interior  compared to that of a rocky body as a function of magma ocean radius. This is the analagous quantity shown  in Fig. 1 of \citet{Peale1979} and Fig. 3 of \citet{seligman2024tidal}. The approximate equilibrium melt radius is where the derivative of this function is zero. This calculation yields a melt radius of $\xi\sim0.6-0.9$. Therefore, if this melting mechanism is operating, we would predict that a significant portion of the interior is melted.

%% For this sample we use BibTeX plus aasjournals.bst to generate the
%% the bibliography. The sample631.bib file was populated from ADS. To
%% get the citations to show in the compiled file do the following:
%%
%% pdflatex sample631.tex
%% bibtext sample631
%% pdflatex sample631.tex
%% pdflatex sample631.tex

\bibliography{sample631}{}
\bibliographystyle{aasjournal}

%% This command is needed to show the entire author+affiliation list when
%% the collaboration and author truncation commands are used.  It has to
%% go at the end of the manuscript.
%\allauthors

%% Include this line if you are using the \added, \replaced, \deleted
%% commands to see a summary list of all changes at the end of the article.
%\listofchanges

\end{document}